%% file: main.tex
\RequirePackage{fix-cm}
\documentclass{article}
\usepackage{microtype}
\usepackage{graphicx}
\usepackage{textcomp}
\usepackage{xcolor}
\usepackage{url}
\usepackage{hyperref}
\usepackage[accepted]{mlsys2025}
\makeatletter
\renewcommand{\Notice@String}{Preprint.}
\makeatother
\usepackage{algorithm}
\usepackage{algorithmic}

\input{preamble}

\input{macros}
\mlsystitlerunning{TileCache}
\begin{document}
\twocolumn[
\mlsystitle{TileCache: Serving Multi-Round Diffusion Image Editing with Compressed Activation Caches}
\begin{mlsysauthorlist}
\mlsysauthor{Yuyang Tian}{uw}
\mlsysauthor{Desen Sun}{uw}
\mlsysauthor{Shuncheng Jie}{uw}
\mlsysauthor{Lishan Yang}{gmu}
\mlsysauthor{Sihang Liu}{uw}
\end{mlsysauthorlist}
\mlsysaffiliation{uw}{Cheriton School of Computer Science, University of Waterloo, Waterloo, ON, Canada}
\mlsysaffiliation{gmu}{Department of Computer Science, George Mason University, Fairfax, VA, USA}
\mlsyscorrespondingauthor{Sihang Liu}{sihangliu@uwaterloo.ca}
\vskip 0.3in
\input{sec/0_abstract}
]
\printAffiliationsAndNotice{}
\hypersetup{pdftitle={TileCache: Serving Multi-Round Diffusion Image Editing with Compressed Activation Caches},
  pdfauthor={Yuyang Tian, Desen Sun, Shuncheng Jie, Lishan Yang, Sihang Liu},
  pdfsubject={arXiv preprint}}

\input{sec/1_intro}
\input{sec/2_background}
\input{sec/3_design}
\input{sec/4_eval}
\input{sec/5_diss}
\input{sec/6_conclusion}

\bibliographystyle{mlsys2025}
\bibliography{bib/references,bib/misc,bib/ml,bib/sys}

\clearpage
\appendix
\input{sec/A_appendix}
\end{document}

%% file: preamble.tex
\usepackage{amsmath}
\usepackage{amssymb}
\usepackage{xspace}
\usepackage{multicol}
\usepackage{multirow}
\usepackage{booktabs}
\usepackage{cleveref}
\usepackage{enumitem}
\usepackage{amsmath}
\usepackage{makecell}
\usepackage{textcomp}
\usepackage{pifont}

\usepackage{array}
\newcolumntype{L}[1]{>{\raggedright\let\newline\\\arraybackslash\hspace{0pt}}m{#1}}
\newcolumntype{C}[1]{>{\centering\let\newline\\\arraybackslash\hspace{0pt}}m{#1}}
\newcolumntype{R}[1]{>{\raggedleft\let\newline\\\arraybackslash\hspace{0pt}}m{#1}}

\usepackage{pifont}

\usepackage{pgfplots}
\usepgfplotslibrary{fillbetween}
\usepackage{tikz}
\usetikzlibrary{matrix,positioning}
\usetikzlibrary{external}
\usepackage{pgfplots}
\usepackage{pgfplotstable}
\usetikzlibrary{pgfplots.groupplots}
\usetikzlibrary{arrows}
\usetikzlibrary{patterns}
\usetikzlibrary{positioning}
\usetikzlibrary{decorations.pathreplacing}
\usetikzlibrary{shapes.arrows}
\usetikzlibrary{shapes.geometric,shapes.misc}
\usetikzlibrary{pgfplots.groupplots}
\pgfplotsset{compat=newest}
\pgfkeys{/pgf/number format/.cd,1000 sep={}}

\pgfplotsset{
    discard if/.style 2 args={
        x filter/.code={
            \edef\tempa{\thisrow{#1}}
            \edef\tempb{#2}
            \ifx\tempa\tempb
                
            \fi
        }
    },
    discard if not/.style 2 args={
        x filter/.code={
            \edef\tempa{\thisrow{#1}}
            \edef\tempb{#2}
            \ifx\tempa\tempb
            \else
                
            \fi
        }
    }
}

\makeatletter%
\newcommand\resetstackedplots{%
\pgfplots@stacked@isfirstplottrue
}
\makeatother

\usetikzlibrary{spy}
\usetikzlibrary{calc}
\usetikzlibrary{patterns,backgrounds}
\pgfdeclarelayer{foreground}
\pgfsetlayers{background,main,foreground}

\usepgfplotslibrary{statistics}

\usepackage[skins]{tcolorbox}

\newcounter{takeawaycounter}
\crefname{takeawaycounter}{Takeaway}{Takeaways}

\newtcolorbox[use counter=takeawaycounter]{takeawaybox}[2]{
  enhanced, width=\hsize,left=1pt,right=1pt,top=1pt,bottom=1pt,colback=black!10!white,boxrule=1pt,colframe=black!30!white,
  label type=takeawaycounter,
  label = #1,
  title = \textcolor{black}{\bf Takeaway~\arabic{takeawaycounter}~#2}
}

\usepackage{needspace}
\usepackage{float}
\usepackage{flafter}
\usepgfplotslibrary{groupplots}
\input{fig/eval_plot_style}
\input{fig/att3x_styles}

%% file: fig/eval_plot_style.tex
\definecolor{evalpink}{HTML}{E0218A}
\definecolor{evalorange}{HTML}{D68B3D}
\definecolor{evalgreen}{HTML}{278A70}
\pgfplotsset{evalaxis/.style={
  font=\sffamily\fontsize{7}{8}\selectfont,
  tick label style={font=\sffamily\fontsize{7}{8}\selectfont},
  label style={font=\sffamily\fontsize{7}{8}\selectfont},
  title style={font=\sffamily\fontsize{7}{8}\selectfont},
  axis line style={black!65},tick style={black!65},
  ymajorgrids=true,grid style={dashed,black!15},
  legend style={draw=none,fill=none,font=\sffamily\fontsize{7}{8}\selectfont},
  scaled ticks=false,clip=false,
}}

%% file: fig/att3x_styles.tex
\colorlet{evalgrey}{black!55}
\definecolor{evalpurple}{HTML}{7B4F9E}
\definecolor{evalbrown}{HTML}{8C6D3F}
\definecolor{evalred}{HTML}{B8433E}
\definecolor{evalgold}{HTML}{C79A1C}
\pgfplotsset{
  e2e line/.style={thick,mark size=1.4pt,unbounded coords=jump},
  arm nocache/.style={e2e line,evalgrey,densely dashed,mark=o,mark options={solid}},
  arm flashps/.style={e2e line,evalorange,dash dot,mark=triangle*,mark options={solid}},
  arm flashpsdct/.style={e2e line,evalpurple,mark=diamond*},
  arm fifo/.style={e2e line,evalbrown,densely dotted,mark=pentagon*,mark options={solid}},
  arm rr/.style={e2e line,evalgold,dash pattern=on 4pt off 1.5pt,mark=asterisk,mark options={solid}},
  arm sjf/.style={e2e line,evalpink,dashed,mark=x,mark options={solid}},
  arm edf/.style={e2e line,evalred,loosely dashed,mark=+,mark options={solid}},
  arm sys/.style={e2e line,evalgreen,very thick,mark=square*},
  att axis/.style={ymin=50,ymax=100,ytick={50,60,...,100},ylabel={SLO attainment (\%)}},
  fair axis/.style={ymin=0.4,ymax=1.0,ytick={0.4,0.6,0.8,1.0},ylabel={Fairness $F=\min S/\max S$}},
}
\newcommand{\targetline}[2]{\draw[dashed,black!65] (axis cs:#1,90) -- (axis cs:#2,90);}

\pgfplotsset{
  fair bar/.style={ybar,draw opacity=0.7,fill opacity=0.32,line width=0.3pt,unbounded coords=discard},
  bars axis/.style={axis y line*=right,axis x line=none,ymajorgrids=false,xlabel={},title={},ymin=0,ymax=1,ytick={0,0.25,0.5,0.75,1}},
  lines axis/.style={axis y line*=left},
  title below/.style={title style={at={(0.5,0)},anchor=north,yshift=-26pt}},
}
\newcommand{\fairbarplot}[3][rps]{\addplot[fair bar,fill=#2,draw=#2] table[x=#1,y=fair,col sep=comma]{#3};}
\newcommand{\fairtitle}[1]{\node[rotate=90,anchor=north,font=\sffamily\fontsize{7}{8}\selectfont] at ($(#1.east)+(0.23in,0)$) {Fairness $F$};}
\newcommand{\fairlegend}{\addlegendimage{area legend,fill=white,draw=black!55}\addlegendentry{Fairness $F$}}

%% file: sec/0_abstract.tex
\begin{abstract}
Multi-round diffusion image editing can reduce computation by reusing intermediate activations, but retaining these activations across editing rounds introduces substantial memory and bandwidth overheads.
Compression can reduce these overheads, but efficient cache construction is challenging: compressing only after editing requires buffering over 100 GB of activations per request.
Moreover, the compressed cache must support selective reuse, as future masked edits require only a subset of the cached activations.
It also exposes a scheduling challenge, as cached edits are faster than initial generations, causing naive policies to severely delay either initial generations or editing requests.
In this paper, we aim to address both challenges through compression-informed serving.
We develop TileCache, a serving system for multi-round diffusion editing. It compresses reusable activations and incorporates a scheduler that balances both edits and initial generation requests.
Compared with the state-of-the-art image editing baseline,
TileCache sustains over 2$\times$ the request arrival rate at 90\% SLO attainment. It completes edits 2.4--4.5$\times$ and initial generations 1.3--2.2$\times$ faster than the baseline, while preserving image quality.

\end{abstract}

%% file: sec/1_intro.tex
\section{Introduction}
\label{sec:intro}

Diffusion models have become an important tool for interactive image creation~\citep{adobe-genfill,midjourney-editor}, allowing users to generate images and refine them through successive edits~\citep{flux2klein,qwenimageedit2511}. In each round, users provide an instruction and a mask specifying the region to modify. The diffusion model then edits only the masked region according to the instruction while preserving consistency between the unmasked regions and the input image. As interactive workflows become increasingly prevalent, efficient edit serving systems are essential for delivering responsive services despite the substantial computational overhead of diffusion models.

To efficiently serve image editing tasks, prior studies have proposed leveraging mask-aware activation reuse to reduce computation \citep{sige,fisedit,flashps}.
This technique recomputes activations only for the masked region while reusing cached activations for the
unmasked region, thereby avoiding repeated processing of unchanged image regions across multiple editing turns.
Despite the computation efficiency, activation reuse in turn requires substantial storage to maintain intermediate activations across editing rounds. For instance, caching activations for FLUX.2 base at $1024\times1024$ resolution over 50 denoising steps requires 107~GB per session (\S\ref{sec:reuse}). Maintaining these caches limits the sessions that fit in memory and introduces additional data transfer when users return for subsequent edits.

Cache compression makes activation reuse more efficient by reducing its memory footprint and data transfer~\citep{kivi,cachegen}, which enables the serving system to retain reusable states across more sessions while reducing data transfer overhead, translating computational savings into end-to-end latency reductions.
However, these optimizations create substantial latency disparities between masked edits with reusable caches and requests without cache reuse.
Such heterogeneity can bias scheduling decisions toward faster requests, potentially starving requests with higher computational demands.
Therefore, reducing cache size alone does not necessarily improve serving quality. Efficiently serving multi-round editing requires not only compressing cached activations but also carefully scheduling requests to balance waiting time and execution interference.

However, achieving this goal poses challenges in both cache compression and request scheduling.
Prior studies have shown that diffusion model activations exhibit similarity across both blocks and denoising steps \cite{adacache,diffcachesurvey}, suggesting opportunities for multi-dimensional compression. However, naively analyzing similarity after all activations are computed incurs prohibitive memory overhead, as retaining these activations throughout a single editing request can require over 100 GB of storage.
Furthermore, since subsequent requests reuse cached activations only in unmasked regions, the compressed representation must support composable reuse, as the regions modified by future edits are unknown in advance. Therefore, the first challenge is how to design an efficient and composable cache compression mechanism to manage the historical cache.
For the scheduler, considering only data transfer and computation latency is insufficient, as the substantial memory footprint of cached activations can lead to out-of-memory errors. Moreover, interference among concurrent requests further complicates scheduling decisions. Designing an effective request placement policy that jointly accounts for latency, memory constraints, and inter-request interference remains challenging.

This work introduces \textbf{TileCache},\footnote{We will open-source TileCache upon paper acceptance.} a serving system for multi-round diffusion image editing that combines cache compression with compression-informed scheduling. To address the first challenge, TileCache develops a compression method that exploits redundancy along two dimensions: denoising steps and transformer blocks. Across steps, TileCache maintains a running sum and a few projections for each block, updating them as new activations are generated. This allows raw activations to be discarded immediately after each step, avoiding the need to retain all intermediate activations.
Across blocks, TileCache compresses each block as a residual relative to the reconstructed previous block, exploiting the similarity between adjacent blocks. The {\it key insight} is that sums and projections are linear: the residual's sums and projections can be obtained by subtraction, without storing the residual activations.
After generation, TileCache processes the blocks sequentially and compresses their residuals using these sums and projections, without accessing the raw activations (\S\ref{sec:compression}).

To address the second challenge, TileCache develops a compression-informed scheduler that accounts for cache memory requirements, execution times, and interference between concurrent requests, all of which depend on the compressed representation. The scheduler first identifies GPUs with sufficient memory to accommodate an incoming request's model activations and cache buffers. For each feasible placement, it estimates the completion time of the incoming request and the additional delay imposed on requests already running on the GPU. Based on these estimates, TileCache determines whether the request should wait or be colocated. Before colocating a new request on a busy GPU, the scheduler checks whether each running request can finish within a given multiple of its single-request latency. The balance between initial generations and edits changes with the request mix and GPU load. Every few minutes, the scheduler compares the slowdown of the two classes and adjusts the latency budget and delay penalty to give higher priority to the class that is falling behind (\S\ref{sec:scheduler}).

We evaluate TileCache on three diffusion models: FLUX.2 base, distilled FLUX.2, and Qwen-Image-Edit.
On a $4\times$ H100 platform and 70\% edit-request ratio based on prior work~\citep{katz,flashps}, TileCache sustains over $2\times$ the arrival rate of the baseline at 90\% attainment and, completes edits 2.4--4.5$\times$ and initial generations 1.3--2.2$\times$ faster. Against the baseline equipped with the same compressed caches, it sustains a 14\% higher arrival rate on H100, and at that load it slows edits and initial generations alike, improving fairness by 32\% (\S\ref{sec:eval_e2e}--\ref{sec:eval_sensitivity}).

We summarize the contributions of this work as follows:
\begin{itemize}[leftmargin=*, noitemsep]
\item We develop a streaming two-dimensional compression method that builds compressed activation caches without storing all intermediate activations. The resulting caches support selective reconstruction and foreground updates during multi-round editing (\S\ref{sec:compression}).
\item We design a compression-informed scheduler that considers cache memory requirements, execution times, and interference when placing and colocating requests. It balances initial generations and edits by adjusting its delay budget and penalties based on each request class's observed latency slowdown (\S\ref{sec:scheduler}).
\item We implement these techniques in TileCache and evaluate its serving performance, compression quality, and individual components across different workloads and GPU platforms. We also validate the simulator against real GPU executions (\S\ref{sec:evaluation}).
\end{itemize}

%% file: sec/2_background.tex
\section{Background and Motivation}
\label{sec:motivation}

\subsection{Activation Reuse and Cache Overhead}
\label{sec:reuse}

Multi-round image editing allows users to refine an image through editing requests. A typical session starts with an initial image generation, followed by edits that modify selected regions of the image. Each edit specifies an instruction and a mask indicating which parts of the image should change.

Existing work has shown that image editing can be accelerated by reusing intermediate activations from previous rounds~\citep{sige,fisedit,flashps}. Instead of recomputing the entire image, mask-aware reuse recomputes tokens in the masked area while retrieving cached activations for tokens in the unmasked area. A cache stores residual-stream outputs from transformer blocks across denoising steps, with separate outputs for each guidance branch when applicable.

\noindent\textbf{Cache footprint and transfer costs.}
Activation reuse introduces substantial storage and transfer overhead. At $1024\times1024$ resolution, an uncompressed cache occupies 107.4~GB for FLUX.2~\citep{flux2klein} and 87.3~GB for Qwen-Image-Edit-2511~\citep{qwenimageedit2511}. A 400~GB host cache can therefore hold only three to four sessions, even before accounting for other memory usage.

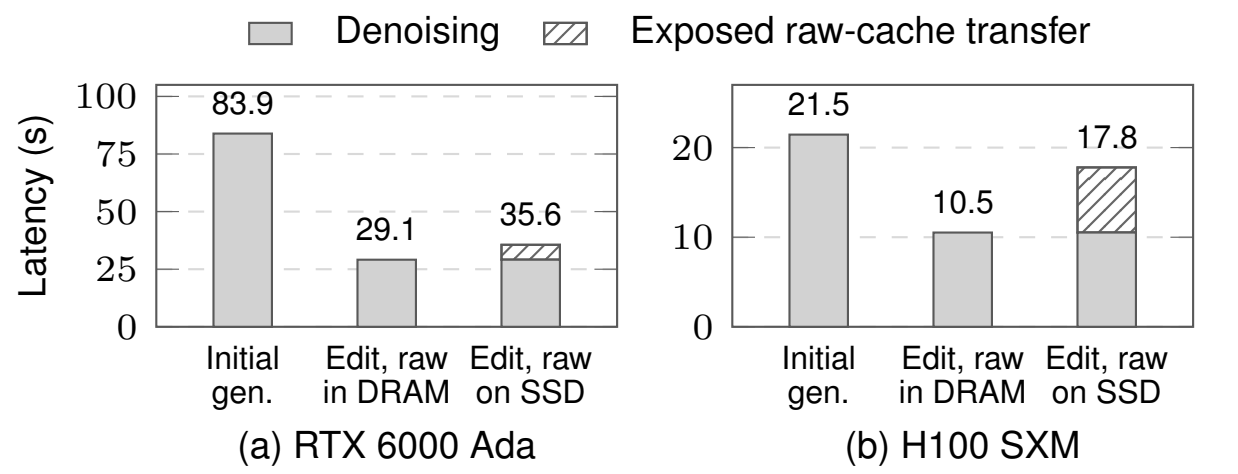
\begin{figure}[t]
\centering
\input{fig/figM1_latency.tikz}
\caption{Activation reuse reduces edit latency, but cache transfer introduces overhead.  Latency comparison for FLUX.2 at $1024\times1024$ with 5.9\% masked tokens.}
\label{fig:motivation_latency}
\end{figure}

Figure~\ref{fig:motivation_latency} shows both the benefits of activation reuse and the overhead of cache transfers. Our measurements of FLUX.2 show that reusing activations from host memory reduces edit latency from 83.9\,s to 29.1\,s on an RTX 6000 Ada GPU and from 21.5\,s to 10.5\,s on an H100. Retrieving the cache from SSD increases the latency to 35.6\,s and 17.8\,s, including 6.5\,s and 7.3\,s of exposed transfer time, respectively. The faster H100 hides less of the transfer, making this overhead larger relative to the edit itself. These costs motivate cache compression. Compression, however, only lowers the cost of each edit in isolation. The observed latency also depends on when and where the edit runs, which we examine next.

\subsection{Scheduling Trade-offs}
\label{sec:scheduling_motivation}

Compression also changes where an edit can run. An edit can share a GPU with a running request only if its cache fits in the remaining GPU memory: an uncompressed FLUX.2 cache (107~GB) cannot reside on an 80~GB H100, whereas a $25\times$ compressed package (4.3~GB) can. Compressed caches thus allow edits to run without waiting for an idle GPU, raising the question of when an edit should run and whether it should share a GPU.

This question is difficult because the two request classes use GPU resources differently. Initial generations execute the full diffusion process and produce reusable activations, while subsequent edits perform less computation but require access to cached data. An edit waiting behind an initial generation may spend more time in the queue than it needs for execution. Giving edits higher priority can reduce their waiting time, but may substantially delay initial generations.

\begin{figure}[t]
\centering
\input{fig/figM2_tradeoff.tikz}
\caption{Slowdowns due to FCFS and SJF scheduling.
}
\label{fig:motivation_tradeoff}
\end{figure}
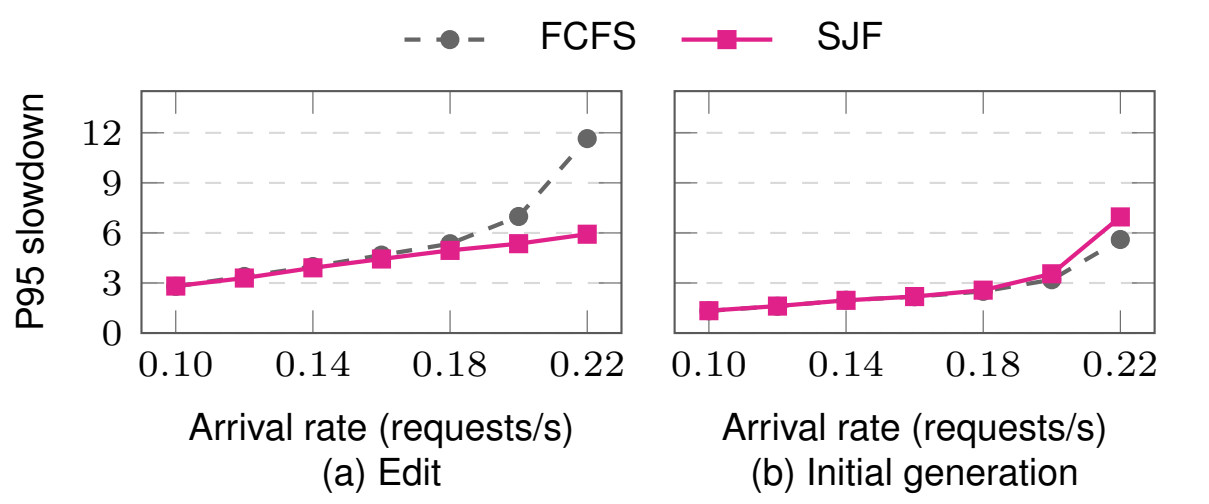

Figure~\ref{fig:motivation_tradeoff} illustrates this trade-off under increasing request rates. We report latency as \emph{slowdown}~\citep{harcholbalter2013}, a request's response latency divided by its single-request latency. We compare first-come-first-served (FCFS) scheduling with colocation against shortest-job-first (SJF) that tends to prioritize shorter editing jobs, both using the same compressed caches. At 0.22 requests/s, edit P95 response latency under FCFS reaches 11.7 times its single-request latency. SJF reduces this ratio to 5.9, but increases initial-generation P95 latency from 5.6 to 7.0 times its single-request latency.

Sharing a GPU does not remove this trade-off. FCFS already allows colocation, but edit latency remains high.
On H100 GPUs, a batched step with two colocated requests takes 93--97\% of their combined single-request step latencies, leaving little room for throughput gains.
Colocating an edit can reduce its waiting time, but the additional GPU work delays requests already running. The scheduler must therefore weigh an edit's waiting time against the delay it would impose, and both depend on the compressed cache: its size determines whether it can stay on the GPU and how much data must be transferred before the edit can start.

\subsection{Design Challenges}
\label{sec:challenges}

The above observations highlight two challenges in serving multi-round image editing workloads: compressing large activation caches without sacrificing efficient reuse, and scheduling requests whose execution and data-access costs depend on the compressed cache.

\noindent\textbf{Challenge 1: Streaming compression with selective access.}
Compression must reduce cache storage and transfer costs without sacrificing image quality or introducing significant memory and reconstruction overhead. Cached activations are redundant across denoising steps, which produce similar activations for the same token~\citep{deepcache,teacache}, and across adjacent transformer blocks, which share a residual stream~\citep{minicache}. Retaining all raw activations during compression incurs substantial storage overhead, making it necessary to exploit both forms of redundancy in a streaming manner.

Since diffusion models generate activations sequentially, compressing earlier activations limits their availability for subsequent steps, requiring efficient algorithms to preserve compression quality.
And for transformer blocks, predicting each block's activations from the reconstructed previous block ensures consistent decoding, but stepwise compression requires the complete denoising trajectory, so the previous block's reconstructed activations are not available when the current block's activations are generated. Combining the two without retaining raw activations is the core difficulty.

Compression must also support the access pattern of multi-round editing. Reconstructing the full cache across all blocks and denoising steps at once would consume substantial GPU memory. Therefore, the compressed cache must support on-demand reconstruction of activations in the unmasked area and updates to activations in the masked area for subsequent editing rounds.

\noindent\textbf{Challenge 2: Scheduling generations and edits.}
As shown in Figure~\ref{fig:motivation_tradeoff}, FCFS and SJF each lead to high latency for one request class. This trade-off appears even at a moderate edit ratio. When edits account for 50\% of requests, prioritizing them can substantially delay initial generations, while FCFS leaves edits waiting behind long-running requests. Thus, the scheduler must adjust its priorities based on the observed latency of each class rather than using a fixed policy.

The scheduler must also decide when to schedule an edit and whether to colocate it with a running request. These decisions depend on the edit's expected waiting time, the remaining work of running requests, and the additional delay caused by colocation. GPU memory further constrains placement decisions, as the model, running requests, and incoming edit must share the available memory, and both an edit's memory need and the delay it causes depend on the size and location of its compressed cache. The compressed cache determines an edit's memory requirements, cache loading time, and per-step reconstruction work, which affect placement feasibility and colocation interference.

Cache transfers also affect scheduling. After an initial generation, the newly built cache must be transferred out of GPU memory before that memory can be released. An edit must load cached activations and write back regenerated activations in the masked area for subsequent rounds. These transfers share bandwidth and can delay cache availability or GPU memory release. An edit may therefore be unable to start even when GPU compute is available.

%% file: fig/figM1_latency.tikz
\pgfplotsset{
  m1 panel/.style={evalaxis,scale only axis,width=1.2in,height=0.63in,anchor=south west,
    ybar stacked,bar width=11pt,xmin=-0.6,xmax=2.6,ymin=0,xtick={0,1,2},
    xticklabels={{Initial\\gen.},{Edit, raw\\in DRAM},{Edit, raw\\on SSD}},
    x tick label style={align=center,font=\sffamily\fontsize{6}{6.5}\selectfont},x tick style={draw=none},
    ylabel shift=-3pt,title style={at={(0.5,0)},anchor=north,yshift=-22pt}},
  m1 denoise/.style={draw=black!65,fill=gray!35},
  m1 transfer/.style={draw=black!65,fill=white,postaction={pattern=north east lines,pattern color=black!60}},
  m1 totals/.style={scale only axis,width=1.2in,height=0.63in,anchor=south west,xmin=-0.6,xmax=2.6,ymin=0,axis lines=none,clip=false,
    nodes near coords,point meta=explicit symbolic,every node near coord/.append style={font=\sffamily\fontsize{6}{6.5}\selectfont,anchor=south}}}
\newcommand{\mpanel}[5]{%
  \begin{axis}[m1 panel,at={(#1,0)},ymax=#3,ytick={#4},#5]
  \addplot[m1 denoise]  table[x expr=\coordindex,y=denoise_s,col sep=comma,comment chars={\#}]{fig/figM1_latency_raw_#2.csv};
  \addplot[m1 transfer] table[x expr=\coordindex,y=transfer_s,col sep=comma,comment chars={\#}]{fig/figM1_latency_raw_#2.csv};
  \end{axis}
  \begin{axis}[m1 totals,at={(#1,0)},ymax=#3]
  \addplot[only marks,mark=none] table[x expr=\coordindex,y=total_s,meta=total_s,col sep=comma,comment chars={\#}]{fig/figM1_latency_raw_#2.csv};
  \end{axis}}
\begin{tikzpicture}
\mpanel{0}{ada}{105}{0,25,50,75,100}{title={(a) RTX 6000 Ada},ylabel={Latency (s)},
  legend to name=m1legend,legend columns=2,legend cell align=left,legend style={column sep=6pt},
  legend image code/.code={\draw[#1] (0pt,-2pt) rectangle (8pt,2pt);}}
\mpanel{1.2in+0.3in}{h100}{27}{0,10,20}{title={(b) H100 SXM}}
\begin{axis}[m1 panel,at={(0,0)},ymax=105,axis lines=none,clip=false,legend to name=m1legend2,legend columns=2,legend cell align=left,
  legend style={column sep=6pt},legend image code/.code={\draw[#1] (0pt,-2pt) rectangle (8pt,2pt);}]
\addlegendimage{m1 denoise}\addlegendentry{Denoising}
\addlegendimage{m1 transfer}\addlegendentry{Exposed raw-cache transfer}
\end{axis}
\node[anchor=south,inner sep=0] at (1.35in,0.63in+0.03in) {\pgfplotslegendfromname{m1legend2}};
\end{tikzpicture}

%% file: fig/figM2_tradeoff.tikz
\begin{tikzpicture}
\begin{groupplot}[group style={group size=2 by 1,horizontal sep=10pt,y descriptions at=edge left},
 evalaxis,scale only axis,width=1.25in,height=0.63in,
 xmin=0.09,xmax=0.23,xtick={0.10,0.14,0.18,0.22},xticklabel style={/pgf/number format/fixed,/pgf/number format/precision=2,/pgf/number format/fixed zerofill},xlabel={Arrival rate (requests/s)},
 ymin=0,ymax=14.5,ytick={0,3,6,9,12},
 every axis plot/.append style={line width=0.8pt,mark size=1.4pt},
 title style={at={(0.5,0)},anchor=north,yshift=-26pt},
 legend columns=3,legend cell align=left,legend style={column sep=6pt},
]
\nextgroupplot[title={(a) Edit},ylabel={P95 slowdown},legend to name=m2legend]
\addplot[black!60,dashed,mark=*,mark options={solid,fill=black!60}]
 table[x=request_rate,y=edit_p95_over_solo,col sep=comma,comment chars={\#}]{fig/figM2_tradeoff_fcfs.csv};
\addlegendentry{FCFS}
\addplot[evalpink,mark=square*,mark options={fill=evalpink}]
 table[x=request_rate,y=edit_p95_over_solo,col sep=comma,comment chars={\#}]{fig/figM2_tradeoff_edf1000share.csv};
\addlegendentry{SJF}
\nextgroupplot[title={(b) Initial generation}]
\addplot[black!60,dashed,mark=*,mark options={solid,fill=black!60}]
 table[x=request_rate,y=firstgen_p95_over_solo,col sep=comma,comment chars={\#}]{fig/figM2_tradeoff_fcfs.csv};
\addplot[evalpink,mark=square*,mark options={fill=evalpink}]
 table[x=request_rate,y=firstgen_p95_over_solo,col sep=comma,comment chars={\#}]{fig/figM2_tradeoff_edf1000share.csv};
\end{groupplot}
\node[anchor=south,inner sep=0] at ($(group c1r1.north)!0.5!(group c2r1.north)+(0,0.03in)$) {\pgfplotslegendfromname{m2legend}};
\end{tikzpicture}

%% file: sec/3_design.tex
\section{TileCache Design}
\label{sec:design}

\subsection{Overview}
\label{sec:overview}

\begin{figure}[t]
\noindent
\includegraphics[width=\columnwidth]{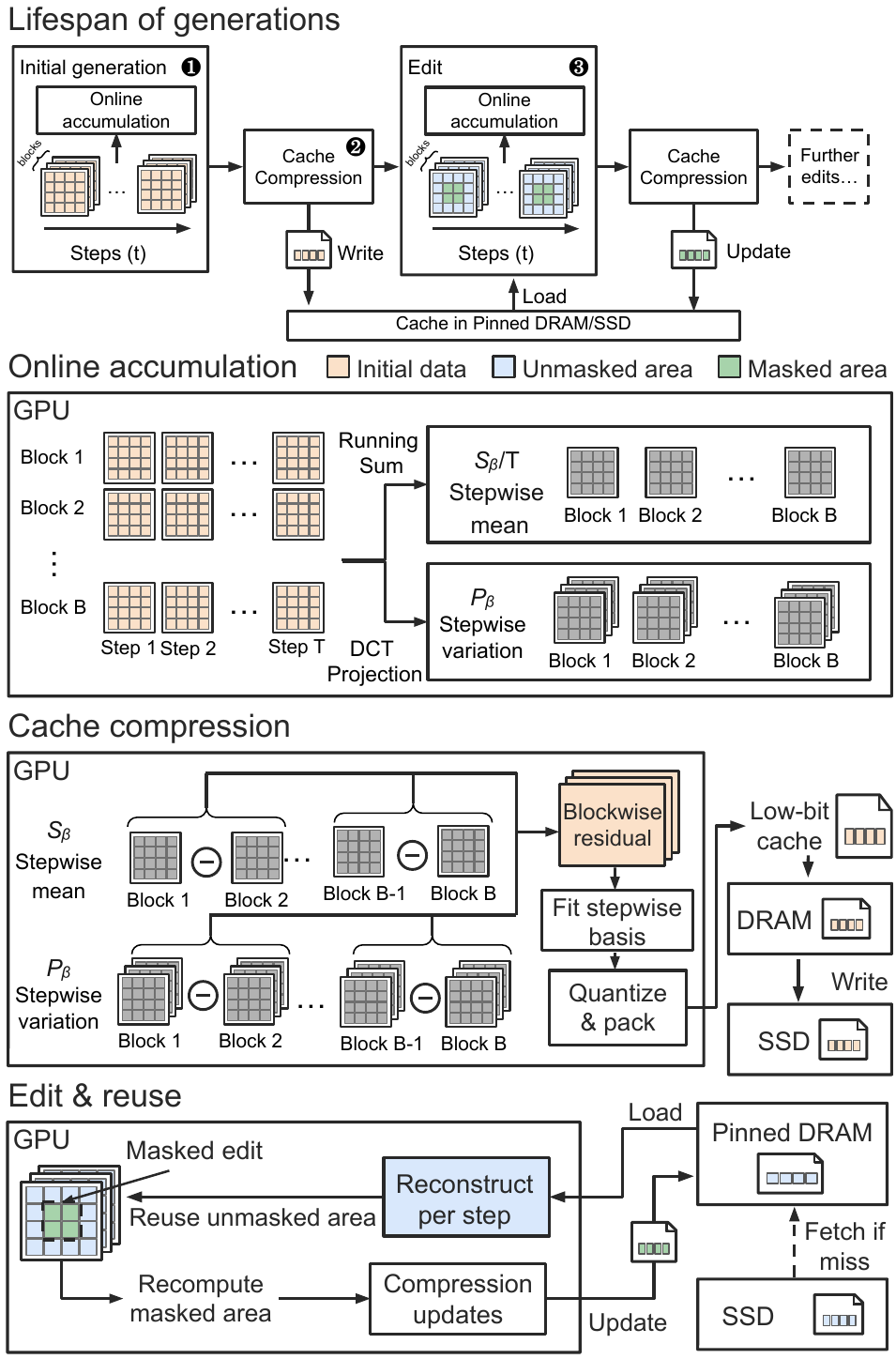}
\caption{Overview of TileCache in multi-round editing.
}
\label{fig:overview}
\end{figure}

\begin{algorithm}[t]
\caption{Compression-Informed Scheduling}
\label{alg:sched}
\begingroup
\fontsize{7.5}{8.8}\selectfont
\makeatletter\ALC@noendtrue\makeatother
\renewcommand{\algorithmicrequire}{\textbf{Scope:}}
\renewcommand{\algorithmicensure}{\textbf{State:}}
\renewcommand{\algorithmiccomment}[1]{\hfill\textit{// #1}}
\begin{algorithmic}[1]
\REQUIRE Incoming edit $r$; all GPUs are busy.
\ENSURE Delay budget $\tau_{\mathrm{eff}}$
    and class weight $w$.

\IF{the latency-feedback window has completed}
    \STATE Compute class P95 slowdowns from requests
        completed in the window
    \IF{at least one class has sufficient samples}
        \STATE Set $\tau_{\mathrm{eff}}$ to the larger of
            $\tau$ and the highest available P95 slowdown
    \ENDIF
    \IF{both classes have sufficient samples}
        \STATE Adjust $w$ using the bounded ratio of
            initial-generation to edit P95 slowdowns
    \ENDIF
\ENDIF

\STATE \textsc{Wait}.impact $\leftarrow$
    single-request latency $+$ estimated waiting time
\STATE $\mathit{options} \leftarrow \{\textsc{Wait}\}$
\FOR{each busy GPU $g$}
    \STATE $\mathit{cache} \leftarrow \textsc{FitCache}(r,g)$
        \COMMENT{GPU first; host streaming}
    \STATE \textbf{if} $\mathit{cache}$ is infeasible
        \textbf{then continue}
    \FOR{each $\mathit{mode}$ in \{shared, pausing\}}
        \STATE $\mathit{option} \leftarrow
            (g,\mathit{cache},\mathit{mode})$
        \STATE Estimate running-request latencies
            for $\mathit{option}$ from profiles
        \IF{any running request's estimated slowdown exceeds
            $\tau_{\mathrm{eff}}$ (Eq.~\ref{eq:sched_budget})}
            \STATE \textbf{continue}
        \ENDIF
        \STATE $\mathit{option}.\mathit{impact} \leftarrow
            \textsc{LatencyImpact}(r,\mathit{option},w)$ (Eq.~\ref{eq:sched_cost})
        \STATE Add $\mathit{option}$ to $\mathit{options}$
    \ENDFOR
\ENDFOR
\STATE \textbf{return} the lowest-impact option in $\mathit{options}$
\end{algorithmic}
\endgroup
\end{algorithm}

Figure~\ref{fig:overview} shows TileCache's architecture and cache lifecycle across an image-editing session. The system consists of a scheduler, GPU workers, and a compressor. Each request includes a session ID and a prompt; edit requests also include a mask specifying the region to modify. The scheduler uses the compressed cache's memory requirements and access costs to guide request placement, with class-level latency feedback helping balance latency between initial generations and edits. The cache lifecycle has three stages:

(1) During initial generation, a worker generates the full image. The compressor accumulates compact summaries of each block's activations on the GPU as denoising proceeds, without retaining raw activations from all steps
(\S\ref{sec:compression}).

(2) Once generation completes, the compressor uses these summaries to build a compressed cache package on the GPU. The package is transferred to pinned host memory for reuse and written to SSD for longer-term storage, releasing the GPU memory occupied by the cache. (\S\ref{sec:compression})

(3) For a subsequent edit, TileCache retrieves the cache and schedules the request (\S\ref{sec:scheduler}). The worker reuses cached activations for the unmasked region, recomputes masked activations, and updates the cache for future edits.

\subsection{Two-Dimensional Activation Compression}
\label{sec:compression}
An editing cache contains activations collected across multiple denoising steps and transformer blocks. TileCache exploits redundancy along both dimensions through stepwise and blockwise compression. It accumulates sums and stepwise projections during generation, then uses them to compress residuals between adjacent transformer blocks without accessing the raw cache (Figure~\ref{fig:overview}). We organize cached activations into spatial rows, each containing all channels of one horizontal row of image tokens across the cached denoising steps. Each row serves as the unit of compression and reconstruction, allowing partial edits without decompressing the entire cache.

\noindent\textbf{Online accumulation and blockwise compression.}
To capture redundancy across denoising steps without retaining the full activation sequence, TileCache accumulates a running sum and stepwise projections as activations arrive. The running sum captures the mean activation across denoising steps, while the projections capture variations around that mean. TileCache uses a fixed DCT dictionary~\citep{dct1974} that provides predefined patterns across denoising steps, allowing these projections to be updated at each step without accessing earlier activations. For each transformer block $b$, we denote the accumulated sum and DCT projections as $S_b$ and $P_b$, respectively. The accumulated sums and projections provide the inputs for subsequent blockwise residual compression and basis fitting without retaining the raw activation sequence. This limits GPU memory use during initial generations.

Adjacent transformer blocks often produce similar activations, making their differences easier to compress than the activations themselves. TileCache therefore compresses the first transformer block directly and represents each subsequent block as the difference between its activations and the reconstructed activations of the preceding transformer block. Let $X_b$ and $\widehat X_b$ denote the original and reconstructed activation sequences of transformer block $b$, respectively, and let $D_b=X_b-\widehat X_{b-1}$, with $\widehat X_0=0$. Because sums and projections are linear, TileCache computes the residual's sums and projections by subtracting those of the reconstructed previous block from those accumulated for the current block, without storing raw residuals:
\begin{equation}
\begin{aligned}
S(D_b) &= S_b-S(\widehat X_{b-1}),\\
P(D_b) &= P_b-P(\widehat X_{b-1}).
\end{aligned}
\label{eq:compression}
\end{equation}
where $S(\cdot)$ and $P(\cdot)$ denote the sum and DCT projections across denoising steps, respectively. After compressing a residual, TileCache reconstructs the current block one step at a time, updates the sums and projections needed for the next block, and discards each reconstructed step immediately afterward.

After generation, TileCache uses the accumulated stepwise projections to adapt the fixed DCT dictionary into a data-dependent stepwise basis for quantization. For the first block, it computes the mean and eigendecomposes the Gram matrix of the stepwise projections to obtain a rotation. Subsequent blocks apply the same procedure to their blockwise residuals. The compressor then assigns bit widths to the mean and coefficients under a fixed budget per image row using a greedy Gaussian distortion model~\citep{huang1963}. The distortion model weights mean error by the number of cached denoising steps and each coefficient's error by 1. Some coefficient slots may receive zero bits. For each mean or coefficient, TileCache uses a shared bit width across channels but computes quantization offsets and step sizes separately for each channel. The quantized values are then bit-packed into the cache package. Package size affects the memory and transfer costs of subsequent edits.

\noindent\textbf{Reconstruction and partial edits.}
During an edit, the decompressor reconstructs the rows containing required unmasked tokens block by block, by adding the decoded residual to the preceding block's reconstructed activations.
The worker recomputes tokens in the masked area and updates their cached activations for subsequent editing rounds, while keeping the remaining activations compressed.
Each row carries a version tag identifying its basis and quantization parameters.
This way, rows compressed in different rounds can coexist, avoiding the overhead of fully reconstructing and recompressing the cache after each edit.
Updating only modified rows also reduces write-back traffic.

\subsection{Compression-Informed Scheduling}
\label{sec:scheduler}

\noindent\textbf{Placement by latency impact.}
\label{sec:sched_pricing}
Waiting and interference are two main sources of additional latency that can cause SLO misses. Colocating an edit can reduce its waiting time, but may delay requests already running on the GPU. Conversely, avoiding colocation can leave edits waiting long enough to miss their latency targets. Therefore, TileCache estimates the \emph{latency impact} of each placement, defined as the incoming request's response time plus the additional delay it imposes on running requests, and selects the placement with the lowest impact.

When a GPU is idle, an incoming request runs on it immediately. Otherwise, TileCache decides whether the request should wait for a GPU or, for an edit, be colocated on a busy GPU. To support colocation, TileCache adopts FlashPS's step-level continuous batching~\citep{flashps}, using one of two execution modes. In shared execution, colocated requests run together in batched steps. In pausing, initial generations are temporarily suspended while the incoming edit executes, and other edits can continue running. The scheduler evaluates both modes for each feasible colocation. Algorithm~\ref{alg:sched} summarizes the scheduling process for an incoming edit when all GPUs are busy.

The compressed cache's memory requirements, staging time, and per-step execution costs determine which colocations are feasible and their latency and interference costs (\S\ref{sec:compression}).
Each placement must first fit in GPU memory alongside the model weights and the running requests. Initial generations require model activations and online accumulation buffers, which replace the activations from all denoising steps.
Edits additionally require compressed caches, reconstruction buffers, and accumulation buffers for updated activations.
TileCache keeps compressed caches on the GPU when memory permits and streams them from host memory otherwise. Their smaller footprint creates more opportunities for GPU-side colocation, and the package size and location set the transfer and reconstruction time of each edit, as estimated below.

For each feasible colocation, TileCache estimates the incoming edit's latency from cache staging ($T_{\mathrm{stage}}$), execution ($T_{\mathrm{exec}}$), and cache updates ($T_{\mathrm{update}}$), using profiled execution and transfer costs.
Cache staging prepares the compressed package for execution, with its cost depending on the cache size, location, and exposed transfer overhead. Execution includes reconstruction in the unmasked area and recomputation in the masked area, while row-level updates limit recompression and write-back to modified rows. TileCache also estimates the total additional delays to running edits ($D_e$) and initial generations ($D_c$).
Under shared execution, TileCache estimates the additional delay to each running request from the incoming edit's profiled per-step execution cost and the number of denoising steps over which the two requests overlap.
Under pausing, an initial generation incurs the full pause duration, while delays to concurrently running edits are estimated using the same overlap-based model.
A running initial generation must also complete GPU-side compression before releasing its buffers. Delaying it can therefore postpone GPU availability for queued initial generations. TileCache increases the weight of such delays based on the number of queued initial generations $q_c$, normalized by the total number of GPUs $G$. An adaptive weight $w$ further adjusts this weight based on the observed latency imbalance between the two request classes, as described below. The resulting latency impact of colocating edit $r$ on GPU $g$ is
\begin{equation}
\begin{aligned}
C_{\mathrm{join}}(r,g)
&= T_{\mathrm{stage}}+T_{\mathrm{exec}}+T_{\mathrm{update}}\\
&\quad+D_e+w\left(1+\frac{q_c}{G}\right)D_c.
\end{aligned}
\label{eq:sched_cost}
\end{equation}

Waiting has latency impact equal to the request's single-request latency $s_r$ plus the estimated time until a GPU becomes available. The scheduler compares this impact with those of colocation and selects the lowest-impact action satisfying the delay constraints described below. By accounting for both waiting and interference, TileCache reduces unnecessary waiting without ignoring the delays imposed on running requests.

\noindent\textbf{Latency-feedback balancing.}
\label{sec:sched_balance}
Latency impact aggregates the delays of all affected requests, so selecting the lowest-impact placement may still cause excessive delays to individual requests or repeatedly impose more delay on one request class. TileCache therefore incorporates class-level latency feedback to adjust the colocation delay budget and the relative weight of interference.

To prevent a low-impact placement from excessively delaying individual requests, TileCache applies a delay budget before allowing colocation. Since initial generations and edits have different single-request latencies, the same additional delay can result in different relative slowdowns. The budget therefore limits each running request's estimated slowdown, consistent with the SLO normalization (\S\ref{sec:eval_setup}). The scheduler estimates each request's completion time using its remaining steps and profiled execution costs under the selected execution mode. A base threshold ($\tau$) controls how much slowdown is allowed. Under high load, a fixed delay budget can reject colocation even when running requests have already accumulated substantial delays. This can leave more incoming edits waiting until they miss their SLOs. Adapting the budget allows the scheduler to consider these placements while still accounting for their interference delays. The scheduler periodically computes the P95 slowdown for each request class from requests completed in the preceding window, requiring at least two samples for a class to be included. TileCache takes the largest available P95 slowdown as $m$ to adjust the effective delay budget.

The delay budget restricts colocation but does not determine how the scheduler balances the two request classes among feasible placements. TileCache therefore adjusts the weight $w$ in Eq.~\ref{eq:sched_cost} using a bounded ratio of the P95 slowdowns for initial generations ($m_c$) and edits ($m_e$). Larger slowdowns for initial generations increase the weight of delaying them, whereas larger slowdowns for edits reduce this weight and encourage colocation. This feedback adjusts how colocation interference is distributed between the two classes based on their observed tail latencies. The weight is updated only when both classes have sufficient samples.

For each candidate colocation on GPU $g$, the scheduler checks the running requests ($\mathcal R_g$) using each request's arrival time ($a_j$), estimated completion time ($f_j^{\mathrm{est}}$), and single-request latency ($s_j$). The effective delay budget ($\tau_{\mathrm{eff}}$) is the larger of $\tau$ and $m$. Colocation is allowed only if
\begin{equation}
\max_{j\in\mathcal R_g}
\frac{f_j^{\mathrm{est}}-a_j}{s_j}
\leq\tau_{\mathrm{eff}},
\qquad
\tau_{\mathrm{eff}}=\max(\tau,m).
\label{eq:sched_budget}
\end{equation}
If a window completes fewer than two requests in both classes, the scheduler keeps the effective budget from the previous window.

Impact-based placement can leave initial generations waiting when edits repeatedly occupy newly available GPUs. To prevent excessive waiting, initial generations waiting longer than their single-request latency receive priority on available GPUs, with the longest-waiting request scheduled first. During this time, pending edits can only join busy GPUs.

%% file: sec/4_eval.tex
\section{Evaluation}
\label{sec:evaluation}

\subsection{Setup and Methodology}
\label{sec:eval_setup}

\noindent\textbf{Platforms and implementation.}
We implement TileCache's compression and decompression using PyTorch~\citep{pytorch} and Triton~\citep{triton}. Its serving runtime consists of a coordinator, one worker per GPU, a shared pinned-memory pool, and direct SSD I/O. We deploy TileCache on $4\times$  H100 SXM GPUs (80\,GB each) and $4\times$ RTX 6000 Ada GPUs (48\,GB each).
For serving evaluation, we use a discrete-event simulator that runs the runtime's scheduling code unmodified, calibrated with GPU execution and transfer measurements from each platform. We validate the simulator against GPU execution at 0.13--0.28 requests/s on H100 and 0.04--0.07 requests/s on RTX 6000 Ada. Simulated class-mean latency differs by at most 1.7\%, P95 slowdown by at most 0.05, and 95\% of scheduling decisions match those of the GPU runtime.
The default storage configuration has 400\,GB of host cache and 4\,TB of SSD, maintaining a 10:1 SSD-to-host-cache capacity ratio throughout the cache-size sensitivity experiments.

\noindent\textbf{Models and workloads.}
Unless noted otherwise, serving experiments use FLUX.2 base~\citep{flux2klein} at $1024\times1024$ resolution and 50 denoising steps. We also evaluate distilled FLUX.2 (4 steps) and Qwen-Image-Edit~\citep{qwenimageedit2511} (50 steps). Sessions follow Midjourney iterative-prompt threads~\citep{midjourneythreads}, with Poisson session arrivals~\citep{flashps,thunderserve} and masked areas sampled from the FlashPS editing trace~\citep{flashps}. The default workload contains 70\% edits, following a diffusion serving trace~\citep{katz} and FlashPS~\citep{flashps}. Each run includes a 24-hour warm-up followed by 48 hours of measurement.

\noindent\textbf{Baselines.}
\emph{FlashPS}~\citep{flashps} uses lossless cache reuse and its original scheduler; caches are preloaded, and SSD writes are excluded. \emph{FlashPS+Compression} uses the same compressed caches as TileCache but retains the FlashPS scheduler, isolating the benefit of compression-informed scheduling.

\noindent\textbf{Serving metrics and load.}
\emph{SLO attainment} is the fraction of requests completing within $3\times$ their class's single-request latency (\S\ref{sec:scheduler})~\citep{alpaserve,distserve}. Following DistServe~\citep{distserve}, \emph{goodput} is the highest arrival rate that achieves at least 90\% SLO attainment. We measure fairness between initial generations and edits as $F=\min_i S_i/\max_i S_i$~\citep{flin}, where $S_i$ is class $i$'s mean latency normalized by its single-request latency; $F=1$ indicates equal slowdowns. We sweep arrival rates through and slightly beyond TileCache's goodput. We set the base delay budget to $\tau=2$ and both the latency-feedback interval and window size to 300\,s (\S\ref{sec:sched_balance}). The sensitivity of this delay budget $\tau$ is evaluated in Appendix~A.1.

\noindent\textbf{Image quality.}
We evaluate editing quality on MagicBrush~\citep{magicbrush}, PIE-Bench~\citep{piebench}, and five-round ImgEdit sessions~\citep{imgedit} at $512\times512$ or $496\times496$ resolution. We use CLIP similarity~\citep{clip} to measure how well an edited image matches its text prompt. Peak signal-to-noise ratio (PSNR) measures pixel-level similarity to a reference image, while structural similarity (SSIM)~\citep{ssim} captures similarity in image structure. Fréchet Inception Distance (FID)~\citep{fid} compares the feature distributions of generated and reference images. Higher CLIP, PSNR, and SSIM scores and lower FID indicate better quality. For PSNR, SSIM, and FID, we use the unmodified Diffusers pipeline~\citep{diffusers}, which recomputes all image tokens without activation reuse. We also compare PSNR and SSIM against FlashPS's lossless reuse~\citep{flashps} to identify compression error.

\subsection{End-to-End Serving Performance}
\label{sec:eval_e2e}

\begin{figure}[t]
\centering
\input{fig/fig6_x70_combined.tikz}
\caption{Serving performance on $4\times$ H100 with 70\% edits.}
\label{fig:eval_e2e}
\end{figure}
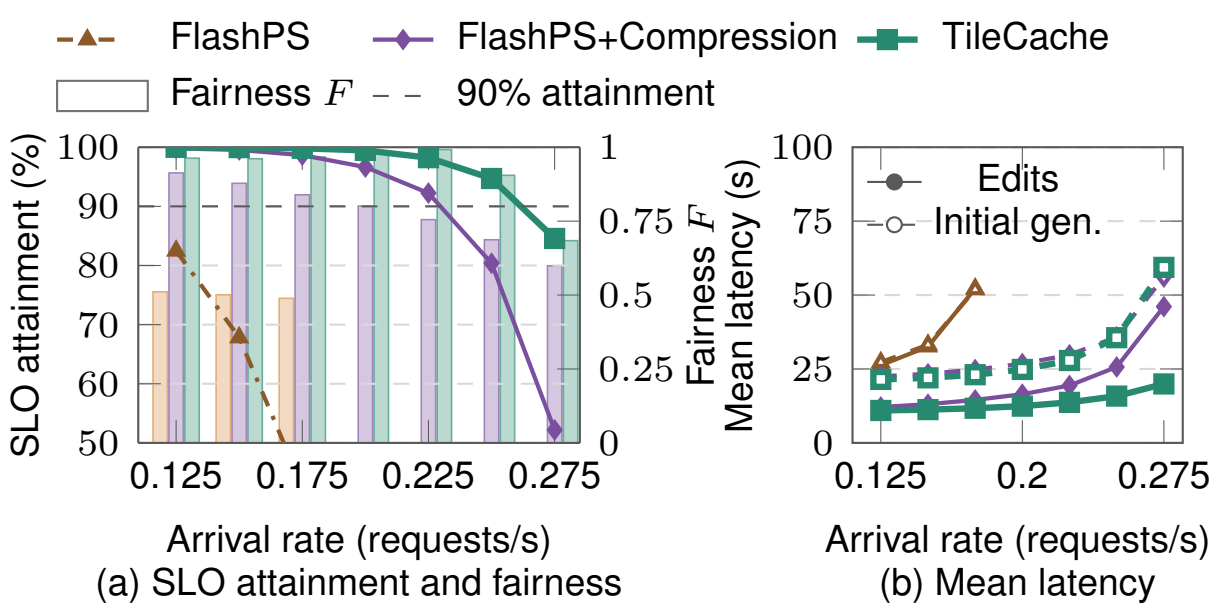
With lossless reuse, FlashPS never reaches the 90\% SLO attainment target, as its large uncompressed caches limit the number of sessions that fit in the 400\,GB host cache (Figure~\ref{fig:eval_e2e}). Its attainment is already 82\% at 0.125 requests/s, less than half of TileCache's goodput. Across the loads it sustains, its edits take 2.4--4.5$\times$ and its initial generations 1.3--2.2$\times$ longer than under TileCache. Giving FlashPS the same compressed caches as TileCache removes this limit. TileCache then achieves a goodput of 0.26 requests/s, 14\% higher than FlashPS+Compression. Since both use the same compressed caches, the improvement comes from compression-informed scheduling.

At 0.25 requests/s, TileCache improves SLO attainment over FlashPS+Compression from 80.4\% to 94.7\% and also the fairness.
It reduces edit slowdown by 38\% on average, with initial-generation slowdown nearly unchanged under FlashPS+Compression.
This improvement comes from comparing the estimated latency costs of colocation and waiting while limiting interference with ongoing requests (\S\ref{sec:scheduler}). TileCache maintains $F\geq0.9$ through 0.25 requests/s, but fairness declines to 0.68 at 0.275 requests/s, beyond its goodput, as more initial generations wait for idle GPUs.

\subsection{Compression Quality}
\label{sec:eval_codec}

\noindent\textbf{(1) Editing quality.}
Despite achieving $23.1$--$26.4\times$ compression for the 50-step models and $4.6\times$ for distilled FLUX.2, TileCache preserves editing quality close to FlashPS's lossless reuse (Table~\ref{tab:e2e_quality}). Across the evaluated datasets, including five-round editing sessions, CLIP scores differ by at most 0.1, while direct output comparisons yield 34.0--46.4\,dB PSNR and 0.973--0.996 SSIM.

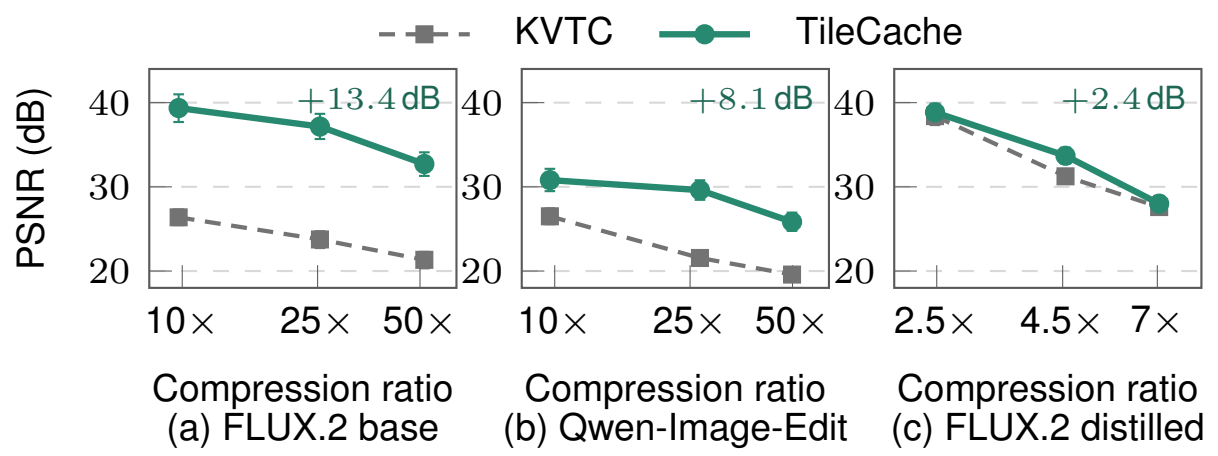
\begin{figure}[t]
\centering
\input{fig/fig7_quality.tikz}
\caption{Masked-area PSNR on MagicBrush.}
\label{fig:eval_quality}
\end{figure}

\noindent\textbf{(2) Comparison with KVTC.}
At comparable compression ratios, TileCache improves masked-area PSNR over KVTC~\citep{kvtc} by 4.3--13.4\,dB on 50-step FLUX.2 base and Qwen-Image-Edit (Figure~\ref{fig:eval_quality}a--b).
On 4-step distilled FLUX.2, TileCache achieves 2.4\,dB higher PSNR at $4.5\times$ compression (Figure~\ref{fig:eval_quality}c).

\input{fig/tab_e2e_quality.tex}

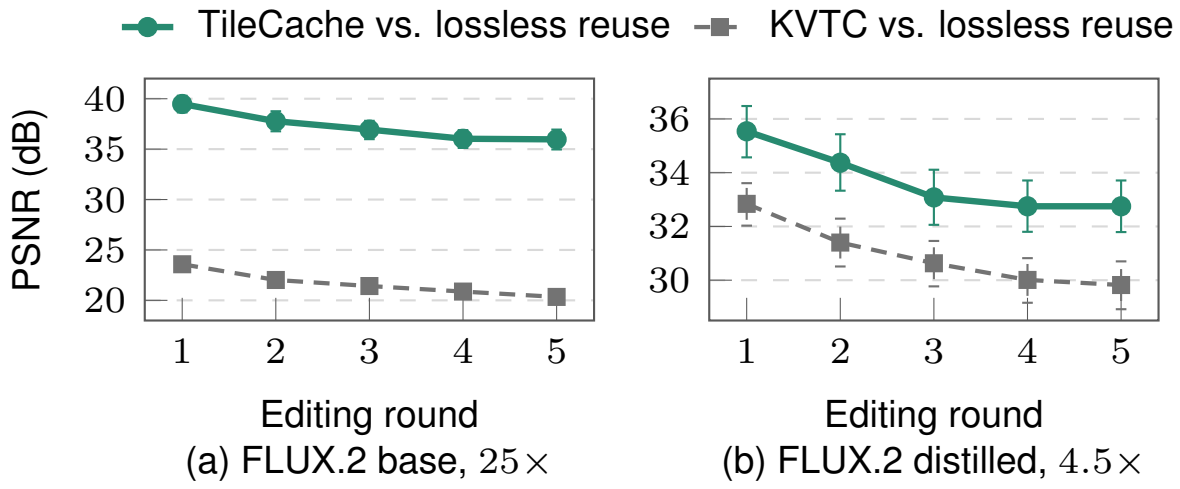
\begin{figure}[t]
\centering
\input{fig/fig_multiround.tikz}
\caption{Masked-area PSNR across 200 ImgEdit sessions.}
\label{fig:eval_multiround}
\end{figure}

\noindent\textbf{(3) Multiple editing rounds.}
The quality advantage persists across repeated edits (Figure~\ref{fig:eval_multiround}). Over 200 five-round ImgEdit sessions, masked-area PSNR relative to lossless reuse decreases from 39.5 to 36.0\,dB for FLUX.2 base and from 35.5 to 32.8\,dB for distilled FLUX.2. TileCache still outperforms KVTC in every round, retaining its quality advantage across successive edits.

\subsection{Ablation Study}
\label{sec:eval_ablation}

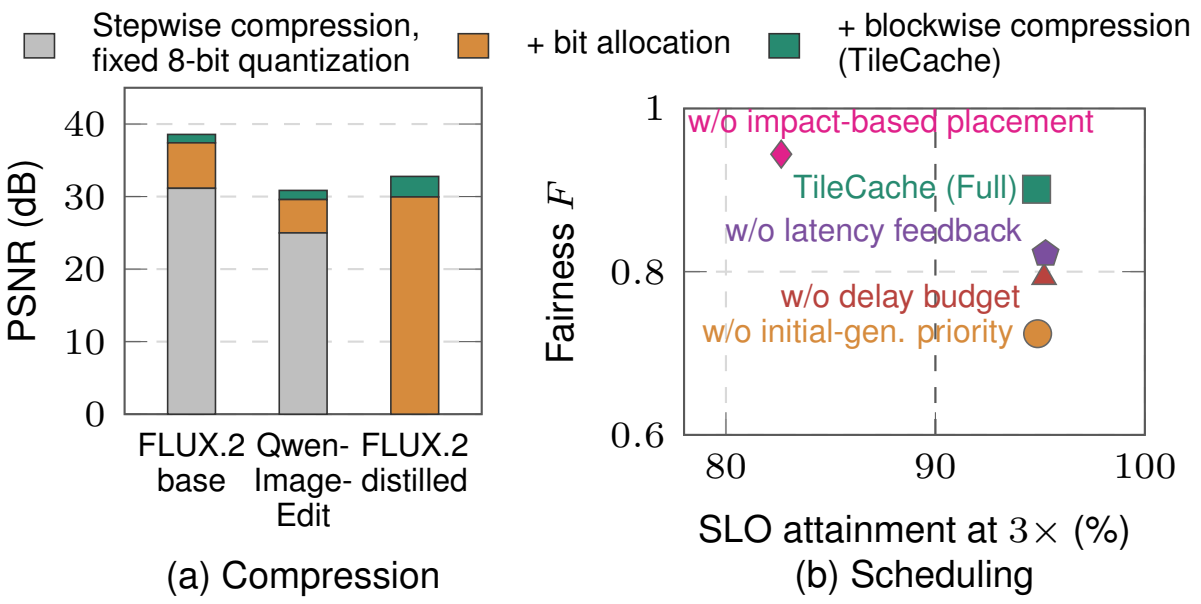
\begin{figure}[t]
\centering
\input{fig/fig_abl_legend.tikz}\\[-4pt]
\begin{minipage}[b]{0.40\columnwidth}\centering\input{fig/fig_abl_codec_stack.tikz}\end{minipage}\hfill
\begin{minipage}[b]{0.58\columnwidth}\centering\input{fig/fig8_ablation_sched.tikz}\end{minipage}
\caption{Ablation study on TileCache components.}
\label{fig:eval_ablation}
\end{figure}

\noindent\textbf{(1) Compression components.}
Compared with stepwise compression using fixed 8-bit quantization, greedy bit allocation improves masked-area PSNR by 6.3\,dB on FLUX.2 base and 4.6\,dB on Qwen-Image-Edit, while blockwise residuals further improve 1.2 and 1.3\,dB, respectively (Figure~\ref{fig:eval_ablation}a).
For distilled FLUX.2, fixed 8-bit quantization cannot achieve the target $4.5\times$ compression ratio. With only 4 denoising steps, stepwise redundancy is limited, making blockwise residuals particularly valuable. The resulting PSNR is 2.8\,dB higher at this ratio and exceeds KVTC by 2.4\,dB. Against KVTC, TileCache achieves its largest gain at $4.5\times$, as shown in Figure~\ref{fig:eval_quality}c.
At $2.5\times$, both compressors exceed 38\,dB, leaving little room for improvement; at $7\times$, the advantage narrows to 0.4\,dB under a tight bit budget. Appendix~A.2 reports 6- and 8-step distilled versions.

\noindent\textbf{(2) Scheduler mechanisms.}
Impact-based placement is critical to SLO attainment (Figure~\ref{fig:eval_ablation}b). Without it, edits join busy GPUs whenever memory permits, increasing interference and reducing attainment from 94.8\% to 82.6\% at 0.25 requests/s. Goodput also drops by 11\%. Although fairness increases from 0.90 to 0.94, both classes experience higher slowdowns, showing that fairness alone does not capture serving performance.

Other mechanisms primarily improve fairness. Removing the delay budget, latency feedback, or initial-generation priority reduces fairness from 0.90 to 0.79, 0.82, and 0.72, respectively, with negligible impact on SLO attainment.
Appendix~A.1 further studies the delay budget.

\subsection{Scheduler Comparison and Sensitivity Analysis}
\label{sec:eval_sensitivity}

\begin{figure}[t]
\centering
\input{fig/fig_sens_sched_combined.tikz}
\caption{Scheduling comparison on $4\times$ H100 with 70\% edits.}
\label{fig:eval_sens_sched}
\end{figure}
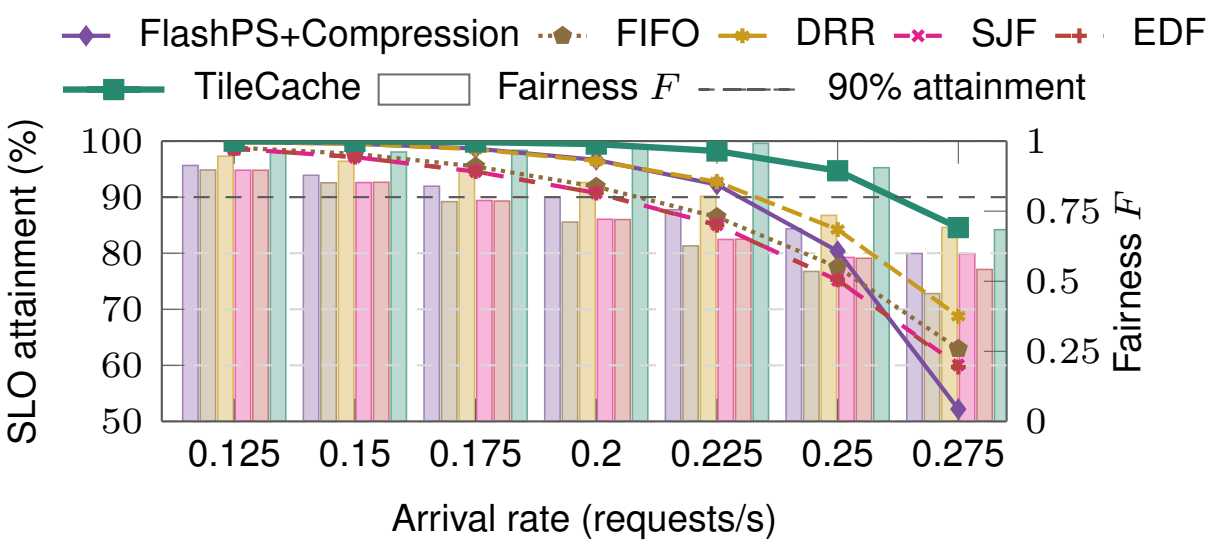

\noindent\textbf{(1) Scheduling policies.}
To isolate scheduling effects, we compare TileCache with FIFO, shortest-job-first (SJF), deficit round robin (DRR)~\citep{drr}, and earliest-deadline-first (EDF), all using the same compressed caches (Figure~\ref{fig:eval_sens_sched}). Each baseline dispatches the head of its queue to an idle GPU, or colocates it on a busy GPU when memory permits; DRR always colocates, while FIFO, SJF, and EDF do so only when it beats waiting for a free GPU. EDF sets each request's deadline to its arrival time plus three times its single-request latency (i.e., the SLO). DRR assigns edits twice the GPU-time weight of initial generations, reflecting their 70\% request share and greater sensitivity to waiting. Since edits have roughly half the single-request latency, the same waiting time contributes approximately twice as much to their slowdown.
DRR falls short of TileCache's goodput by 11\%, and SJF and EDF by 23\%.
At 0.25 requests/s, TileCache achieves fairness of 0.91, compared with 0.74 for DRR, 0.59 for SJF, and 0.58 for EDF. FIFO achieves 20\% lower goodput and a fairness of 0.54. All four baselines delay edits more than initial generations, since colocated edits run at the slower batched step rate.
At 0.275 requests/s, TileCache still meets SLO for 85\% of requests, compared with 60--69\% for the baselines.

\begin{figure}[t]
\centering
\input{fig/fig_sens_platform_combined.tikz}
\caption{GPU platform sensitivity with 70\% edits.
}
\label{fig:eval_sens_platform}
\end{figure}
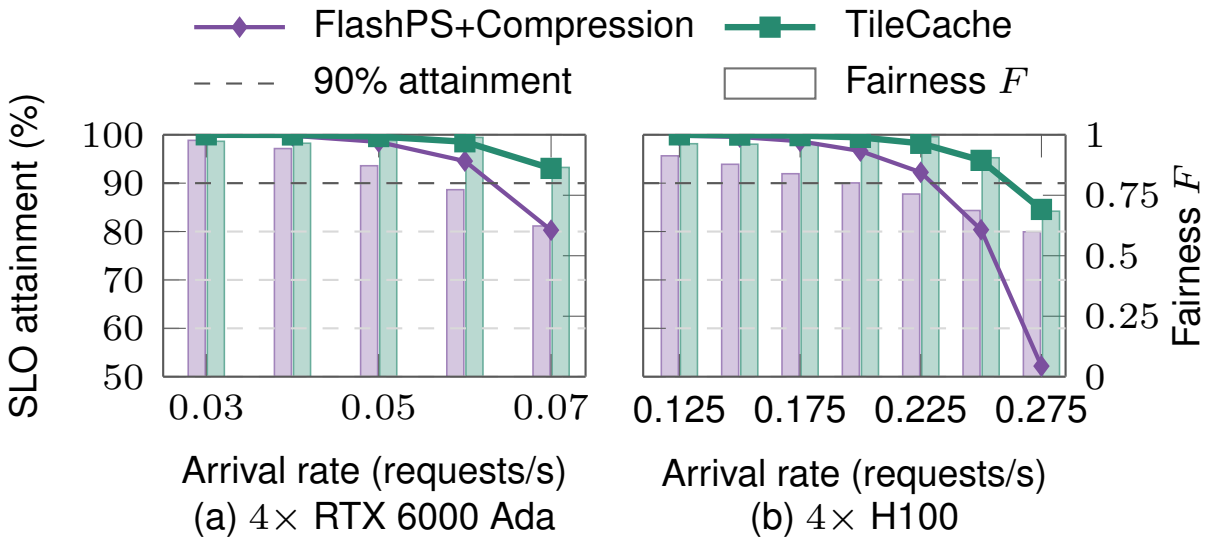

\noindent\textbf{(2) GPU platform.}
TileCache retains its advantage on different GPU platforms (Figure~\ref{fig:eval_sens_platform}). Goodput improves over FlashPS+Compression by 12\% on RTX 6000 Ada and 14\% on H100. The smaller gain on Ada is consistent with its 48\,GB GPU memory, which leaves fewer opportunities for edits to join busy GPUs. At FlashPS+Compression's goodput on each platform, TileCache increases fairness from 0.77 to 0.99 on Ada and from 0.74 to 0.98 on H100.

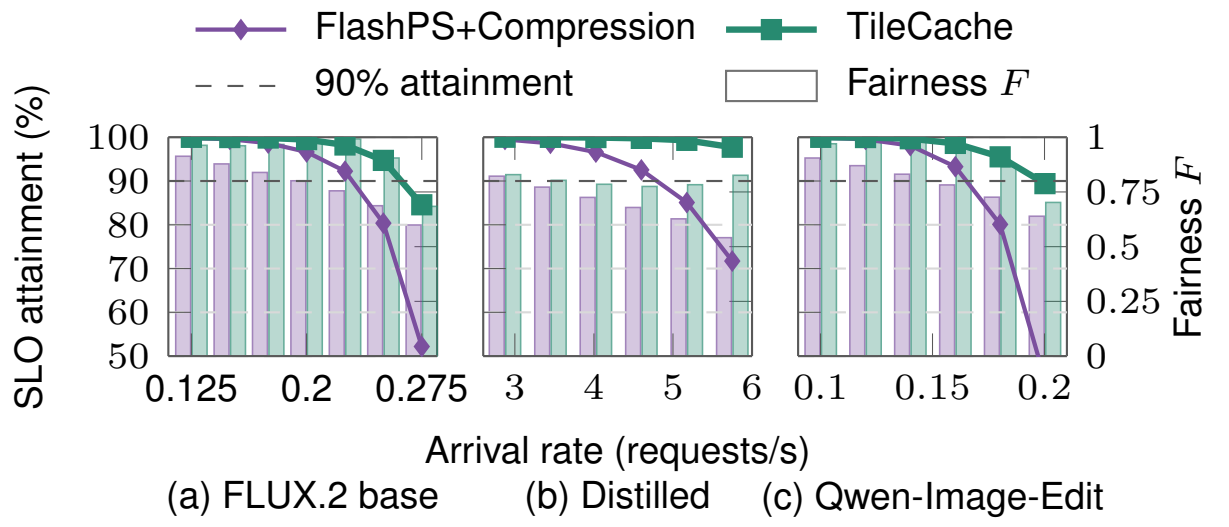
\begin{figure}[t]
\centering
\input{fig/fig_model_combined.tikz}
\caption{Model sensitivity on H100 across
diffusion models.}
\label{fig:eval_model}
\end{figure}

\noindent\textbf{(3) Model.}
The gains extend across models with different denoising lengths (Figure~\ref{fig:eval_model}). TileCache improves goodput over FlashPS+Compression by 14\% on FLUX.2 base, 22\% on 4-step distilled FLUX.2, and 20\% on Qwen-Image-Edit. The gain on Qwen-Image-Edit is due to SSD reads on FlashPS+Compression's critical path following host-cache misses. TileCache also improves fairness by 19--21\% on Qwen-Image-Edit at 0.16--0.18 requests/s and by 14\% on distilled FLUX.2 at 4.6 requests/s.

\begin{figure}[t]
\centering
\input{fig/fig_sens_knobs_combined.tikz}
\caption{Edit-fraction and cache-size sensitivity on $4\times$ H100.}
\label{fig:eval_sens}
\end{figure}
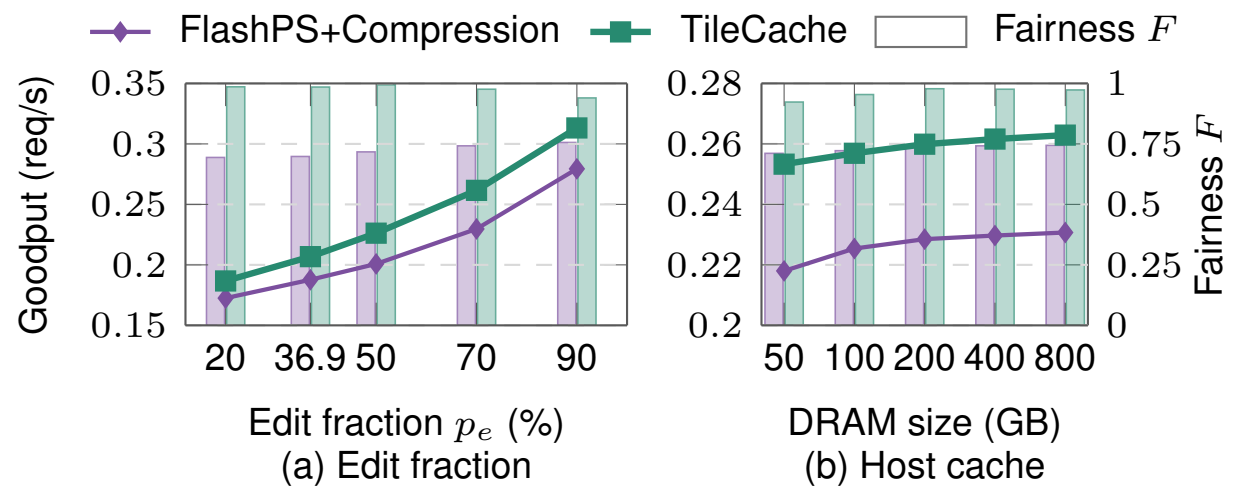

\noindent\textbf{(4) Edit fraction.}
The scheduling benefit persists across mixtures with 20--90\% edits (Figure~\ref{fig:eval_sens}a). TileCache delivers 8\% higher goodput than FlashPS+Compression at 20\% edits and 12--14\% higher goodput at 50--90\% edits, where more incoming edits can join busy GPUs. At the baseline's goodput for each mixture, TileCache maintains fairness of 0.94--0.99, compared with 0.68--0.76 for FlashPS+Compression.

\noindent\textbf{(5) Cache capacity.}
TileCache remains effective even with a small host cache (Figure~\ref{fig:eval_sens}b). As host cache capacity decreases $16\times$ from 800 to 50\,GB, TileCache's cache hit rate falls from 96\% to 32\%, yet its goodput drops by only 4\%. Across this range, TileCache maintains higher goodput and fairness than FlashPS+Compression. Appendix~A.3 breaks down how edits are served at each cache size.

%% file: fig/fig6_x70_combined.tikz
\begin{tikzpicture}
\pgfplotsset{mainpanel/.style={evalaxis,scale only axis,height=0.77in,anchor=south west,
  xmin=0.11,xmax=0.285,restrict x to domain=0:0.275,clip=true,clip marker paths=true,
  xlabel={Arrival rate (requests/s)},
  xticklabel style={/pgf/number format/fixed,/pgf/number format/precision=3},
  xtick={0.125,0.175,0.225,0.275},xticklabels={0.125,0.175,0.225,0.275},
  title below}}
\pgfplotsset{fair bar/.style={ybar,bar width=2.7pt,draw opacity=0.7,fill opacity=0.32,line width=0.3pt,unbounded coords=discard}}
\pgfplotsset{lat edit/.style={solid},lat cold/.style={dashed,mark options={fill=white,solid}}}
\pgfplotsset{arm flashps/.append style={evalorange!65!black}}
\begin{axis}[mainpanel,name=bars,width=1.15in,at={(0,0)},axis y line*=right,axis x line=none,ymajorgrids=false,xlabel={},
  ymin=0,ymax=1,ytick={0,0.25,0.5,0.75,1},ybar=0.3pt]
\addplot[fair bar,fill=evalorange,draw=evalorange] table[x=rps,y=fair,col sep=comma]{fig/att3x/rate_main_flashps.csv};
\addplot[fair bar,fill=evalpurple,draw=evalpurple]     table[x=rps,y=fair,col sep=comma]{fig/att3x/rate_main_flashps-dct.csv};
\addplot[fair bar,fill=evalgreen,draw=evalgreen]   table[x=rps,y=fair,col sep=comma]{fig/att3x/rate_main_sys.csv};
\end{axis}
\node[rotate=90,anchor=north,font=\sffamily\fontsize{7}{8}\selectfont] at ($(bars.east)+(0.23in,0)$) {Fairness $F$};
\begin{axis}[mainpanel,width=1.15in,at={(0,0)},att axis,axis y line*=left,title={(a) SLO attainment and fairness},ylabel shift=-5pt,
  legend to name=mainatt3xlegend,legend cell align=left,legend style={legend columns=3,column sep=2pt,/tikz/every even column/.append style={column sep=1pt}},
  legend image post style={mark size=1.4pt},
  legend image code/.code={\draw[#1,mark repeat=2,mark phase=2] plot coordinates {(0cm,0cm)(0.2cm,0cm)(0.4cm,0cm)};}]
\addplot[arm flashps]    table[x=rps,y=att,col sep=comma]{fig/att3x/rate_main_flashps.csv};
\addplot[arm flashpsdct] table[x=rps,y=att,col sep=comma]{fig/att3x/rate_main_flashps-dct.csv};
\addplot[arm sys]        table[x=rps,y=att,col sep=comma]{fig/att3x/rate_main_sys.csv};
\addlegendentry{FlashPS}\addlegendentry{FlashPS+Compression}\addlegendentry{TileCache}
\addlegendimage{area legend,fill=white,draw=black!55}\addlegendentry{Fairness $F$}
\addlegendimage{dashed,black!65,no markers}\addlegendentry{90\% attainment}
\targetline{0.11}{0.285}
\end{axis}
\begin{axis}[mainpanel,width=0.86in,at={(1.86in,0)},title={(b) Mean latency},
  ymin=0,ymax=100,ytick={0,25,50,75,100},ylabel={Mean latency (s)},ylabel shift=-5pt,xtick={0.125,0.2,0.275},xticklabels={0.125,0.2,0.275},
  legend style={at={(0.03,0.99)},anchor=north west,inner sep=0.5pt,row sep=-2.5pt,fill=none},
  legend image code/.code={\draw[#1,mark repeat=2,mark phase=2] plot coordinates {(0cm,0cm)(0.18cm,0cm)(0.36cm,0cm)};},
  legend image post style={mark size=1.4pt}]
\addlegendimage{black!65,solid,mark=*,mark options={solid}}\addlegendentry{Edits}
\addlegendimage{black!65,dashed,mark=*,mark options={fill=white,solid}}\addlegendentry{Initial gen.}
\addplot[arm flashps,lat edit]    table[x=rps,y=lat_edit,col sep=comma]{fig/att3x/rate_main_flashps.csv};
\addplot[arm flashps,lat cold]    table[x=rps,y=lat_cold,col sep=comma]{fig/att3x/rate_main_flashps.csv};
\addplot[arm flashpsdct,lat edit] table[x=rps,y=lat_edit,col sep=comma]{fig/att3x/rate_main_flashps-dct.csv};
\addplot[arm flashpsdct,lat cold] table[x=rps,y=lat_cold,col sep=comma]{fig/att3x/rate_main_flashps-dct.csv};
\addplot[arm sys,lat edit]        table[x=rps,y=lat_edit,col sep=comma]{fig/att3x/rate_main_sys.csv};
\addplot[arm sys,lat cold]        table[x=rps,y=lat_cold,col sep=comma]{fig/att3x/rate_main_sys.csv};
\end{axis}
\node[anchor=south,inner sep=0] at (1.17in,0.77in+0.03in) {\pgfplotslegendfromname{mainatt3xlegend}};
\end{tikzpicture}

%% file: fig/tab_e2e_quality.tex
\begin{table*}[t]
\centering\scriptsize\setlength{\tabcolsep}{2pt}
\caption{End-to-end edit quality of TileCache and FlashPS}
\label{tab:e2e_quality}
\begin{tabular}{llrrrrrrrrrrrr}
\toprule
 & & & \multicolumn{3}{c}{CLIP $\uparrow$} & \multicolumn{6}{c}{Deviation from the Diffusers edit (no reuse)} & \multicolumn{2}{c}{Error added by compression} \\
\cmidrule(lr){4-6}\cmidrule(lr){7-12}\cmidrule(lr){13-14}
 & & & & & & \multicolumn{2}{c}{PSNR $\uparrow$} & \multicolumn{2}{c}{SSIM $\uparrow$} & \multicolumn{2}{c}{FID $\downarrow$} & \multicolumn{2}{c}{TileCache vs.\ FlashPS edit} \\
\cmidrule(lr){7-8}\cmidrule(lr){9-10}\cmidrule(lr){11-12}\cmidrule(lr){13-14}
Model & Dataset & Ratio & Diffusers & FlashPS & TileCache & FlashPS & TileCache & FlashPS & TileCache & FlashPS & TileCache & PSNR $\uparrow$ & SSIM $\uparrow$ \\
\midrule
FLUX.2 distilled & MagicBrush & 4.6$\times$ & 26.1 & 26.2 & 26.2 & 27.1 & 26.5 & 0.937 & 0.937 & 31.5 & 31.5 & 43.4 & 0.994 \\
 & PIE-Bench & 4.6$\times$ & 26.2 & 26.0 & 26.0 & 21.6 & 21.6 & 0.868 & 0.868 & 60.6 & 60.8 & 38.2 & 0.989 \\
 & ImgEdit (5 rounds) & 4.6$\times$ & 27.6 & 27.8 & 27.8 & 23.9 & 23.9 & 0.911 & 0.911 & 21.4 & 21.5 & 42.2 & 0.993 \\
FLUX.2 base & MagicBrush & 25.1$\times$ & 26.0 & 26.1 & 26.1 & 33.4 & 33.1 & 0.974 & 0.974 & 9.8 & 9.6 & 46.4 & 0.996 \\
 & PIE-Bench & 23.1$\times$ & 25.3 & 25.2 & 25.2 & 29.9 & 29.8 & 0.962 & 0.962 & 20.1 & 20.4 & 43.5 & 0.994 \\
 & ImgEdit (5 rounds) & 24.6$\times$ & 27.7 & 27.7 & 27.7 & 28.7 & 28.6 & 0.951 & 0.951 & 16.2 & 16.2 & 45.3 & 0.996 \\
Qwen-Image-Edit & MagicBrush & 26.4$\times$ & 26.2 & 26.4 & 26.5 & 28.3 & 28.3 & 0.946 & 0.947 & 19.4 & 19.1 & 36.7 & 0.980 \\
 & PIE-Bench & 23.1$\times$ & 25.9 & 25.7 & 25.7 & 33.4 & 33.4 & 0.948 & 0.947 & 26.6 & 26.0 & 40.6 & 0.981 \\
 & ImgEdit (5 rounds) & 25.7$\times$ & 27.8 & 27.9 & 27.9 & 24.1 & 24.1 & 0.903 & 0.903 & 25.6 & 26.0 & 34.0 & 0.973 \\
\bottomrule
\end{tabular}
\end{table*}

%% file: fig/fig7_quality.tikz
\pgfplotstableread[col sep=comma,comment chars={\#}]{fig/fig7_quality_delta.csv}\qdelta
\colorlet{evalgrey}{black!55}
\pgfplotsset{
  q line/.style={thick,mark size=1.3pt,error bars/y dir=both,error bars/y explicit,
                 error bars/error bar style={line width=.4pt},error bars/error mark options={rotate=90,mark size=1pt}},
  q kvtc/.style={q line,evalgrey,densely dashed,mark=square*,mark options={solid}},
  q ours/.style={q line,evalgreen,very thick,mark=*},
}
\newcommand{\qseries}[1]{%
  \addplot[q kvtc] table[x=ratio_x,y=fg,y error plus expr=\thisrow{fg_hi}-\thisrow{fg},
                         y error minus expr=\thisrow{fg}-\thisrow{fg_lo},col sep=comma,comment chars={\#}]{fig/quality/#1_kvtc.csv};
  \addplot[q ours] table[x=ratio_x,y=fg,y error plus expr=\thisrow{fg_hi}-\thisrow{fg},
                         y error minus expr=\thisrow{fg}-\thisrow{fg_lo},col sep=comma,comment chars={\#}]{fig/quality/#1_ours.csv};
}
\newcommand{\qannot}[3]{%
  \pgfplotstableread[col sep=comma,comment chars={\#}]{fig/quality/#1_ours.csv}\qours
  \pgfplotstablegetelem{1}{ratio_x}\of\qours\edef\qx{\pgfplotsretval}%
  \pgfplotstablegetelem{1}{fg}\of\qours\edef\qy{\pgfplotsretval}%
  \pgfplotstablegetelem{#2}{dfg}\of\qdelta\edef\qd{\pgfplotsretval}%
  \node[anchor=north east,font=\sffamily\fontsize{6}{7}\selectfont,evalgreen!75!black,inner sep=1pt]
    at (rel axis cs:0.97,0.95) {$\pgfmathprintnumber[fixed,precision=1,showpos]{\qd}$\,dB};
}
\begin{tikzpicture}
\begin{groupplot}[evalaxis,
  group style={group size=3 by 1,horizontal sep=0.17in,ylabels at=edge left},
  scale only axis,width=0.80in,height=0.57in,
  xmode=log,log ticks with fixed point,xtick pos=left,ytick pos=left,
  ymin=18,ymax=44,ytick={20,30,40},
  ylabel={PSNR (dB)},
  xlabel={Compression ratio},
  every axis title/.append style={at={(0.5,0)},anchor=north,yshift=-26pt},
  legend style={at={(0.5,1.0)},anchor=south,inner ysep=1pt,legend columns=2,column sep=6pt},
  legend image post style={mark size=1.3pt}]
\nextgroupplot[title={(a) FLUX.2 base},xmin=8,xmax=64,xtick={10,25,50},xticklabels={10$\times$,25$\times$,50$\times$}]
\qseries{base_flux2}\qannot{base_flux2}{10}{south west}
\nextgroupplot[title={(b) Qwen-Image-Edit},xmin=8,xmax=64,xtick={10,25,50},xticklabels={10$\times$,25$\times$,50$\times$}]
\qseries{base_qwen}\qannot{base_qwen}{13}{south west}
\addlegendentry{KVTC}
\addlegendentry{TileCache}
\nextgroupplot[title={(c) FLUX.2 distilled},xmin=2.05,xmax=8.6,xtick={2.5,4.5,7},xticklabels={2.5$\times$,4.5$\times$,7$\times$}]
\qseries{distilled_s4}\qannot{distilled_s4}{1}{south west}
\end{groupplot}
\end{tikzpicture}

%% file: fig/fig_multiround.tikz
\colorlet{evalgrey}{black!55}
\pgfplotsset{
  mr line/.style={thick,mark size=1.3pt,error bars/y dir=both,error bars/y explicit,
                  error bars/error bar style={line width=.4pt},error bars/error mark options={rotate=90,mark size=1pt}},
  mr kvtc/.style={mr line,evalgrey,densely dashed,mark=square*,mark options={solid}},
  mr ours/.style={mr line,evalgreen,very thick,mark=*},
  mr panel/.style={evalaxis,scale only axis,width=1.17in,height=0.63in,anchor=south west,
    xmin=0.6,xmax=5.4,xtick={1,2,3,4,5},xlabel={Editing round},xtick pos=left,ytick pos=left,
    title style={at={(0.5,0)},anchor=north,yshift=-26pt}}}
\newcommand{\mrseries}[1]{%
  \addplot[mr ours] table[x=round,y=psnr,y error plus expr=\thisrow{hi}-\thisrow{psnr},y error minus expr=\thisrow{psnr}-\thisrow{lo},
                          col sep=comma,comment chars={\#}]{fig/multiround/#1_ours.csv};
  \addplot[mr kvtc] table[x=round,y=psnr,y error plus expr=\thisrow{hi}-\thisrow{psnr},y error minus expr=\thisrow{psnr}-\thisrow{lo},
                          col sep=comma,comment chars={\#}]{fig/multiround/#1_kvtc.csv};}
\begin{tikzpicture}
\begin{axis}[mr panel,at={(0,0)},title={(a) FLUX.2 base, $25\times$},ymin=18,ymax=42,ytick={20,25,30,35,40},ylabel={PSNR (dB)},
  legend to name=multiroundlegend,legend cell align=left,legend style={legend columns=2,column sep=2pt},legend image post style={mark size=1.3pt},
  legend image code/.code={\draw[#1,mark repeat=2,mark phase=2] plot coordinates {(0cm,0cm)(0.17cm,0cm)(0.34cm,0cm)};}]
\mrseries{base}
\addlegendentry{TileCache vs. lossless reuse}\addlegendentry{KVTC vs. lossless reuse}
\end{axis}
\begin{axis}[mr panel,at={(1.17in+0.3in,0)},title={(b) FLUX.2 distilled, $4.5\times$},ymin=28.5,ymax=37.5,ytick={30,32,34,36}]
\mrseries{dist}
\end{axis}
\node[anchor=south,inner sep=0] at (1.32in,0.63in+0.03in) {\pgfplotslegendfromname{multiroundlegend}};
\end{tikzpicture}

%% file: fig/fig_abl_legend.tikz
\begin{tikzpicture}
\begin{axis}[hide axis,xmin=0,xmax=1,ymin=0,ymax=1,scale only axis,width=1pt,height=1pt,
  legend columns=3,legend cell align=left,
  legend style={at={(0,0)},anchor=south,inner ysep=1pt,draw=none,fill=none,font=\sffamily\fontsize{6}{6.5}\selectfont,nodes={align=left},column sep=5pt},
  legend image code/.code={\draw[#1] (0cm,-0.08cm) rectangle (0.2cm,0.08cm);}]
\addlegendimage{fill=black!25,draw=black!80,line width=0.3pt}\addlegendentry{Stepwise compression,\\fixed 8-bit quantization}
\addlegendimage{fill=evalorange,draw=black!80,line width=0.3pt}\addlegendentry{+ bit allocation}
\addlegendimage{fill=evalgreen,draw=black!80,line width=0.3pt}\addlegendentry{+ blockwise compression\\(TileCache)}
\end{axis}
\end{tikzpicture}

%% file: fig/fig_abl_codec_stack.tikz
\begin{tikzpicture}
\begin{axis}[evalaxis,scale only axis,width=0.93in,height=0.85in,
  ybar stacked,bar width=9pt,xmin=-0.6,xmax=2.6,xtick={0,1,2},xticklabels={FLUX.2\\base,Qwen-\\Image-\\Edit,FLUX.2\\distilled},
  xticklabel style={align=center,font=\sffamily\fontsize{6}{6.5}\selectfont},x tick style={draw=none},
  ymin=0,ymax=45,ytick={0,10,20,30,40},ylabel={PSNR (dB)},ylabel shift=-3pt,
  title={(a) Compression},title style={at={(0.5,0)},anchor=north,yshift=-30pt},
  ]
\addplot[fill=black!25,draw=black!80,line width=0.3pt]  table[x expr=\coordindex,y=stepwise,col sep=comma]{fig/fig_abl_codec_stack_wide.csv};
\addplot[fill=evalorange,draw=black!80,line width=0.3pt] table[x expr=\coordindex,y=bitalloc,col sep=comma]{fig/fig_abl_codec_stack_wide.csv};
\addplot[fill=evalgreen,draw=black!80,line width=0.3pt]  table[x expr=\coordindex,y=blockwise,col sep=comma]{fig/fig_abl_codec_stack_wide.csv};
\end{axis}
\end{tikzpicture}

%% file: fig/fig8_ablation_sched.tikz
\pgfplotsset{abl pt/.style={only marks,mark size=2.6pt,mark options={draw=black!60,line width=0.3pt}}}
\tikzset{abl lab/.style={font=\sffamily\fontsize{6}{6.5}\selectfont,inner sep=0.3pt}}
\begin{tikzpicture}
\begin{axis}[evalaxis,scale only axis,width=1.2in,height=0.85in,title={(b) Scheduling},title below,
  xmin=78,xmax=100,xtick={80,90,100},xlabel={SLO attainment at $3\times$ (\%)},
  ymin=0.6,ymax=1.0,ytick={0.6,0.8,1.0},ylabel={Fairness $F$},
  xmajorgrids=true,clip=false]
\draw[dashed,black!65] (axis cs:90,0.6) -- (axis cs:90,1.0);
\addplot[abl pt,mark=square*,evalgreen]  table[x=att,y=fair,col sep=comma,restrict expr to domain={\thisrow{idx}}{0:0}]{fig/att3x/abl_points.csv}
  node[abl lab,anchor=east,xshift=-3pt] {TileCache (Full)};
\addplot[abl pt,mark=diamond*,evalpink]  table[x=att,y=fair,col sep=comma,restrict expr to domain={\thisrow{idx}}{1:1}]{fig/att3x/abl_points.csv}
  node[abl lab,anchor=south west,xshift=-18pt,yshift=2.5pt] {w/o impact-based placement};
\addplot[abl pt,mark=triangle*,evalred]  table[x=att,y=fair,col sep=comma,restrict expr to domain={\thisrow{idx}}{2:2}]{fig/att3x/abl_points.csv}
  node[abl lab,anchor=east,xshift=-4pt,yshift=-4pt] {w/o delay budget};
\addplot[abl pt,mark=pentagon*,evalpurple] table[x=att,y=fair,col sep=comma,restrict expr to domain={\thisrow{idx}}{3:3}]{fig/att3x/abl_points.csv}
  node[abl lab,anchor=east,xshift=-4pt,yshift=4pt] {w/o latency feedback};
\addplot[abl pt,mark=*,evalorange]       table[x=att,y=fair,col sep=comma,restrict expr to domain={\thisrow{idx}}{4:4}]{fig/att3x/abl_points.csv}
  node[abl lab,anchor=east,xshift=-4pt] {w/o initial-gen. priority};
\end{axis}
\end{tikzpicture}

%% file: fig/fig_sens_sched_combined.tikz
\begin{tikzpicture}
\pgfplotsset{panel/.style={evalaxis,scale only axis,width=2.2in,height=0.73in,anchor=south west,at={(0,0)},
  xmin=0.11,xmax=0.285,restrict x to domain=0:0.275,clip=true,clip marker paths=true,
  xlabel={Arrival rate (requests/s)},xticklabel style={/pgf/number format/fixed,/pgf/number format/precision=3},
  xtick={0.125,0.15,0.175,0.2,0.225,0.25,0.275},xticklabels={0.125,0.15,0.175,0.2,0.225,0.25,0.275},title style={yshift=-3pt}}}
\begin{axis}[panel,bars axis,name=bars,bar width=2.9pt,ybar=0.4pt,legend to name=schedlegrow2,legend columns=-1,legend style={column sep=3pt,/tikz/every even column/.append style={column sep=1pt}},legend image post style={mark size=1.4pt},
  legend image code/.code={\draw[#1,mark repeat=2,mark phase=2] plot coordinates {(0cm,0cm)(0.16cm,0cm)(0.32cm,0cm)};}]
\addlegendimage{arm sys}\addlegendentry{TileCache}
\fairlegend
\addlegendimage{dashed,black!65,no markers}\addlegendentry{90\% attainment}
\fairbarplot{evalpurple}{fig/att3x/rate_sched_flashps-dct.csv}
\fairbarplot{evalbrown}{fig/att3x/rate_sched_fifojoin.csv}
\fairbarplot{evalgold}{fig/att3x/rate_sched_rrjoin.csv}
\fairbarplot{evalpink}{fig/att3x/rate_sched_sjfjoin.csv}
\fairbarplot{evalred}{fig/att3x/rate_sched_edfjoin3.csv}
\fairbarplot{evalgreen}{fig/att3x/rate_sched_sys.csv}
\end{axis}
\fairtitle{bars}
\begin{axis}[panel,att axis,lines axis,legend to name=schedlegrow1,legend columns=-1,legend style={column sep=3pt,/tikz/every even column/.append style={column sep=1pt}},legend image post style={mark size=1.4pt},
  legend image code/.code={\draw[#1,mark repeat=2,mark phase=2] plot coordinates {(0cm,0cm)(0.16cm,0cm)(0.32cm,0cm)};}]
\addplot[arm flashpsdct] table[x=rps,y=att,col sep=comma]{fig/att3x/rate_sched_flashps-dct.csv};
\addplot[arm fifo]       table[x=rps,y=att,col sep=comma]{fig/att3x/rate_sched_fifojoin.csv};
\addplot[arm rr]         table[x=rps,y=att,col sep=comma]{fig/att3x/rate_sched_rrjoin.csv};
\addplot[arm sjf]        table[x=rps,y=att,col sep=comma]{fig/att3x/rate_sched_sjfjoin.csv};
\addplot[arm edf]        table[x=rps,y=att,col sep=comma]{fig/att3x/rate_sched_edfjoin3.csv};
\addplot[arm sys,forget plot] table[x=rps,y=att,col sep=comma]{fig/att3x/rate_sched_sys.csv};
\addlegendentry{FlashPS+Compression}\addlegendentry{FIFO}\addlegendentry{DRR}\addlegendentry{SJF}\addlegendentry{EDF}
\targetline{0.11}{0.285}
\end{axis}
\node[anchor=south west,inner sep=0] (row2) at (-0.31in,0.73in+0.03in) {\pgfplotslegendfromname{schedlegrow2}};
\node[anchor=south west,inner sep=0] at (-0.31in,0.73in+0.03in+0.14in) {\pgfplotslegendfromname{schedlegrow1}};
\end{tikzpicture}

%% file: fig/fig_sens_platform_combined.tikz
\begin{tikzpicture}
\pgfplotsset{panel/.style={evalaxis,scale only axis,width=1.1in,height=0.63in,anchor=south west,clip=true,clip marker paths=true,
  xlabel={Arrival rate (requests/s)},xticklabel style={/pgf/number format/fixed,/pgf/number format/precision=3},title below},
  ada/.style={at={(0,0)},xmin=0.025,xmax=0.074,restrict x to domain=0:0.072,xtick={0.03,0.05,0.07}},
  h100/.style={at={(1.25in,0)},xmin=0.11,xmax=0.285,restrict x to domain=0:0.275,xtick={0.125,0.175,0.225,0.275},xticklabels={0.125,0.175,0.225,0.275}}}
\begin{axis}[panel,ada,bars axis,yticklabels={},bar width=3.2pt,ybar=0.4pt]
\fairbarplot{evalpurple}{fig/att3x/rate_ada_flashps-dct.csv}
\fairbarplot{evalgreen}{fig/att3x/rate_ada_sys.csv}
\end{axis}
\begin{axis}[panel,h100,bars axis,name=bars,bar width=3.2pt,ybar=0.4pt]
\fairbarplot{evalpurple}{fig/att3x/rate_h100_flashps-dct.csv}
\fairbarplot{evalgreen}{fig/att3x/rate_h100_sys.csv}
\end{axis}
\fairtitle{bars}
\begin{axis}[panel,ada,att axis,lines axis,title={(a) $4\times$ RTX 6000 Ada},ylabel={SLO attainment (\%)},legend to name=platatt3xlegend,legend cell align=left,
  legend style={legend columns=2,column sep=3pt},legend image post style={mark size=1.4pt}]
\addplot[arm flashpsdct] table[x=rps,y=att,col sep=comma]{fig/att3x/rate_ada_flashps-dct.csv};
\addplot[arm sys]        table[x=rps,y=att,col sep=comma]{fig/att3x/rate_ada_sys.csv};
\addlegendentry{FlashPS+Compression}\addlegendentry{TileCache}
\addlegendimage{dashed,black!65,no markers}\addlegendentry{90\% attainment}
\fairlegend
\targetline{0.025}{0.074}
\end{axis}
\begin{axis}[panel,h100,att axis,lines axis,title={(b) $4\times$ H100},ylabel={},yticklabels={}]
\addplot[arm flashpsdct] table[x=rps,y=att,col sep=comma]{fig/att3x/rate_h100_flashps-dct.csv};
\addplot[arm sys]        table[x=rps,y=att,col sep=comma]{fig/att3x/rate_h100_sys.csv};
\targetline{0.11}{0.285}
\end{axis}
\node[anchor=south,inner sep=0] at (1.18in,0.63in+0.03in) {\pgfplotslegendfromname{platatt3xlegend}};
\end{tikzpicture}

%% file: fig/fig_model_combined.tikz
\begin{tikzpicture}
\pgfplotsset{panel/.style={evalaxis,scale only axis,width=0.7in,height=0.57in,anchor=south west,clip=true,clip marker paths=true,
  xticklabel style={/pgf/number format/fixed,/pgf/number format/precision=2},title below},
  base/.style={at={(0,0)},xmin=0.11,xmax=0.285,restrict x to domain=0:0.275,xtick={0.125,0.2,0.275},xticklabels={0.125,0.2,0.275}},
  dist/.style={at={(0.82in,0)},xmin=2.6,xmax=6.0,restrict x to domain=0:5.75,xtick={3,4,5,6}},
  qwen/.style={at={(1.64in,0)},xmin=0.09,xmax=0.21,restrict x to domain=0:0.205,xtick={0.1,0.15,0.2}}}
\begin{axis}[panel,base,bars axis,yticklabels={},bar width=2.8pt,ybar=0.3pt]
\fairbarplot{evalpurple}{fig/att3x/rate_h100_flashps-dct.csv}
\fairbarplot{evalgreen}{fig/att3x/rate_h100_sys.csv}
\end{axis}
\begin{axis}[panel,dist,bars axis,yticklabels={},bar width=2.8pt,ybar=0.3pt]
\fairbarplot{evalpurple}{fig/att3x/rate_dist_flashps-dct.csv}
\fairbarplot{evalgreen}{fig/att3x/rate_dist_sys.csv}
\end{axis}
\begin{axis}[panel,qwen,bars axis,name=bars,bar width=2.8pt,ybar=0.3pt]
\fairbarplot{evalpurple}{fig/att3x/rate_qwen_flashps-dct.csv}
\fairbarplot{evalgreen}{fig/att3x/rate_qwen_sys.csv}
\end{axis}
\fairtitle{bars}
\begin{axis}[panel,base,att axis,lines axis,title={(a) FLUX.2 base},ylabel={SLO attainment (\%)},legend to name=modelatt3xlegend,legend cell align=left,
  legend style={legend columns=2,column sep=3pt},legend image post style={mark size=1.4pt}]
\addplot[arm flashpsdct] table[x=rps,y=att,col sep=comma]{fig/att3x/rate_h100_flashps-dct.csv};
\addplot[arm sys]        table[x=rps,y=att,col sep=comma]{fig/att3x/rate_h100_sys.csv};
\addlegendentry{FlashPS+Compression}\addlegendentry{TileCache}
\addlegendimage{dashed,black!65,no markers}\addlegendentry{90\% attainment}
\fairlegend
\targetline{0.11}{0.285}
\end{axis}
\begin{axis}[panel,dist,att axis,lines axis,title={(b) Distilled},ylabel={},yticklabels={},xlabel={Arrival rate (requests/s)}]
\addplot[arm flashpsdct] table[x=rps,y=att,col sep=comma]{fig/att3x/rate_dist_flashps-dct.csv};
\addplot[arm sys]        table[x=rps,y=att,col sep=comma]{fig/att3x/rate_dist_sys.csv};
\targetline{2.6}{6.0}
\end{axis}
\begin{axis}[panel,qwen,att axis,lines axis,title={(c) Qwen-Image-Edit},ylabel={},yticklabels={}]
\addplot[arm flashpsdct] table[x=rps,y=att,col sep=comma]{fig/att3x/rate_qwen_flashps-dct.csv};
\addplot[arm sys]        table[x=rps,y=att,col sep=comma]{fig/att3x/rate_qwen_sys.csv};
\targetline{0.09}{0.21}
\end{axis}
\node[anchor=south,inner sep=0] at (1.17in,0.57in+0.03in) {\pgfplotslegendfromname{modelatt3xlegend}};
\end{tikzpicture}

%% file: fig/fig_sens_knobs_combined.tikz
\begin{tikzpicture}
\pgfplotsset{panel/.style={evalaxis,scale only axis,height=0.63in,anchor=south west,clip=true,clip marker paths=true,title below},
  mix/.style={at={(0,0)},width=1.15in,xmin=12,xmax=100,restrict x to domain=15:100,xtick={20,36.9,50,70,90},xticklabels={20,36.9,50,70,90},xlabel={Edit fraction $p_e$ (\%)}},
  dram/.style={at={(1.5in,0)},width=0.85in,xmin=40,xmax=1000,xtick={50,100,200,400,800},xticklabels={50,100,200,400,800},xlabel={DRAM size (GB)}}}
\begin{axis}[panel,mix,bars axis,yticklabels={},bar width=3.4pt,ybar=0.4pt]
\fairbarplot[pe]{evalpurple}{fig/att3x/knob_mix_flashps-dct.csv}
\fairbarplot[pe]{evalgreen}{fig/att3x/knob_mix_sys.csv}
\end{axis}
\begin{axis}[panel,dram,xmode=log,log basis x=2,bars axis,name=bars,bar width=3.4pt,ybar=0.4pt]
\fairbarplot[gb]{evalpurple}{fig/att3x/knob_dram_flashps-dct.csv}
\fairbarplot[gb]{evalgreen}{fig/att3x/knob_dram_sys.csv}
\end{axis}
\fairtitle{bars}
\begin{axis}[panel,mix,lines axis,title={(a) Edit fraction},ylabel={Goodput (req/s)},ymin=0.15,ymax=0.35,ytick={0.15,0.2,0.25,0.3,0.35},
  legend to name=knobatt3xlegend,legend cell align=left,legend style={legend columns=3,column sep=3pt},legend image post style={mark size=1.4pt},
  legend image code/.code={\draw[#1,mark repeat=2,mark phase=2] plot coordinates {(0cm,0cm)(0.2cm,0cm)(0.4cm,0cm)};}]
\addplot[arm flashpsdct] table[x=pe,y=cap,col sep=comma]{fig/att3x/knob_mix_flashps-dct.csv};
\addplot[arm sys]        table[x=pe,y=cap,col sep=comma]{fig/att3x/knob_mix_sys.csv};
\addlegendentry{FlashPS+Compression}\addlegendentry{TileCache}
\fairlegend
\end{axis}
\begin{axis}[panel,dram,xmode=log,log basis x=2,lines axis,title={(b) Host cache},ylabel={},ymin=0.2,ymax=0.28,ytick={0.2,0.22,0.24,0.26,0.28}]
\addplot[arm flashpsdct] table[x=gb,y=cap,col sep=comma]{fig/att3x/knob_dram_flashps-dct.csv};
\addplot[arm sys]        table[x=gb,y=cap,col sep=comma]{fig/att3x/knob_dram_sys.csv};
\end{axis}
\node[anchor=south,inner sep=0] at (1.18in,0.63in+0.03in) {\pgfplotslegendfromname{knobatt3xlegend}};
\end{tikzpicture}

%% file: sec/5_diss.tex
\section{Related Work}
\label{sec:related}

\textbf{KV caching and compression.} LLM serving systems compress KV state with techniques such as cross-layer prediction, transform coding, and low-bit quantization~\citep{xquantcl,aquakv,kvtc,freqkv,turboquant}, and cache it across requests in host-memory, disk, or GPU/CPU tiers with loads overlapped against compute or streamed in compressed form~\citep{attentionstore,pensieve,cachegen}.
These methods compress completed KV states for reuse by later requests. TileCache instead accumulates compact statistics as DiT block outputs are generated, exploiting redundancy across denoising steps and blocks without retaining the activations from all denoising steps. It also uses the compressed cache's memory and transfer costs to decide whether an edit should join a GPU running an initial generation to achieve better SLO attainment.

\textbf{Diffusion reuse.} Step-level caching reuses or forecasts features across denoising steps within one generation~\citep{deepcache,teacache,taylorseer}, and approximate caching resumes generation from noisy latents cached from similar prompts~\citep{nirvana,flexcache}; neither holds the per-step block outputs that masked editing of a specific image requires. SIGE~\citep{sige} and FISEdit~\citep{fisedit} cache the original image's per-step activations and recompute only edited regions with sparse kernels.
FlashPS~\citep{flashps} brings this into a serving system, caching large, uncompressed block outputs and batching edits at step granularity, but it treats both the cache's size and its creation as outside the serving problem.
TileCache tackles both problems by compressing cached activations to keep the working set in memory and scheduling the initial generations that create caches with the edits that consume them.

\textbf{Scheduling for model serving.} Model serving systems schedule LLM requests at iteration granularity over paged KV memory~\citep{orca,vllm} and exploit predictable execution costs and per-request latency state to meet SLOs~\citep{clockwork,sola}; diffusion serving systems route queries through model cascades or vary per-request parallelism across denoising steps to meet deadlines~\citep{diffserve,tetriserve}.
None of these systems address scheduling for a cache lifecycle which involves the initial generation and subsequent edits that reuse the cache.
TileCache uses the resulting memory and latency costs to compare waiting with colocation. It allows an edit to join a busy GPU only when memory permits and the estimated slowdown of running requests remains within the delay budget.

%% file: sec/6_conclusion.tex
\section{Conclusions}
This paper presents TileCache, a serving system for multi-round diffusion editing. It compresses reusable activations and incorporates a scheduler that balances both edits and initial generation requests.
Streaming two-dimensional compression avoids retaining raw activations, and compression-informed colocation bounds the delay an edit imposes on running requests.
We evaluate TileCache with three diffusion models and demonstrate over $2\times$ the arrival rate of FlashPS at 90\% SLO attainment and 14--22\% more than FlashPS equipped with the same compressed caches, with edits completing 2.4--4.5$\times$ and initial generations 1.3--2.2$\times$ faster, without noticeable quality difference.

%% file: sec/A_appendix.tex
\setcounter{figure}{0}
\renewcommand{\thefigure}{A\arabic{figure}}
\setcounter{table}{0}
\renewcommand{\thetable}{A\arabic{table}}

\section{Additional Results}

\subsection{Delay-Budget Sensitivity}
\label{sec:app_ablation}

The delay budget improves fairness with little impact on goodput (Table~A1). Removing the delay budget allows edits to colocate more freely, but increases goodput by only 2\%. At 0.25 requests/s, fairness falls from 0.91 at $\tau=2$ to 0.86 at $\tau=3$, 0.81 at $\tau=5$, and 0.80 without the budget, as looser budgets allow edits to join busy GPUs and delay initial generations. At 0.275 requests/s, beyond goodput, removing the budget reduces fairness from 0.68 to 0.56. Since $\tau=1.5$ achieves nearly the same fairness as $\tau=2$, we use $\tau=2$, the largest tested budget that maintains $F\geq0.9$ at 0.25 requests/s.

\subsection{Distilled FLUX.2 with More Denoising Steps}
\label{sec:app_quality_distilled}

TileCache's advantage over KVTC grows with the number of denoising steps at higher compression ratios (Figure~A1). At $4.5\times$ compression, it improves masked-area PSNR by 2.4, 3.8, and 4.0\,dB at 4, 6, and 8 steps, respectively. Blockwise residuals further improve PSNR by 2.5--3.0\,dB over stepwise-only compression. At $2.5\times$, both methods achieve 38--41\,dB, with differences below 0.6\,dB. At $7\times$, TileCache's advantage grows from 0.4\,dB at 4 steps to 2.0 and 3.2\,dB at six and eight steps, as longer denoising trajectories provide more opportunities to exploit redundancy across steps.

\subsection{Host-Cache Hit Rates}
\label{sec:app_hitrate}

A smaller host cache shifts more edits from host-memory hits to SSD fetches (Table~A2). As host-cache capacity shrinks from 800\,GB to 50\,GB, with SSD capacity set to ten times host capacity, the host-cache hit rate falls from 96\% to 32\%, the share of edits requiring SSD fetches rises from 3\% to 62\%, and the share whose caches are no longer available on SSD increases from 1.1\% to 5.6\%.

\input{fig/tab_app_tau.tex}

\input{fig/tab_app_hitrate.tex}

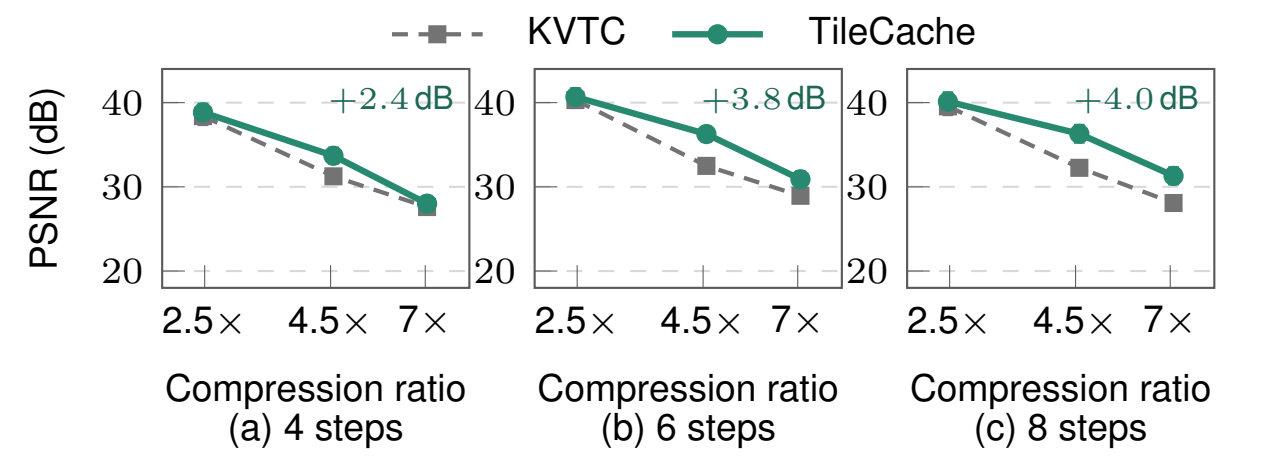
\begin{figure}[H]
\centering
\input{fig/fig7_quality_distilled.tikz}
\caption{Masked-area PSNR on MagicBrush for distilled FLUX.2 at 4, 6, and 8 denoising steps.}
\label{fig:app_quality_distilled}
\end{figure}

%% file: fig/tab_app_tau.tex
\begin{table}[H]
\centering\scriptsize\setlength{\tabcolsep}{2pt}
\caption{Delay-budget sensitivity on $4\times$ H100.}
\label{tab:app_tau}
\begin{tabular}{lrrrrr}
\toprule
 & & \multicolumn{2}{c}{0.25 requests/s} & \multicolumn{2}{c}{0.275 requests/s} \\
\cmidrule(lr){3-4}\cmidrule(lr){5-6}
Delay budget $\tau$ & \makecell[r]{Goodput\\ (\%)} & \makecell[r]{SLO\\ attainment (\%)} & \makecell[r]{Fairness\\ $F$} & \makecell[r]{SLO\\ attainment (\%)} & \makecell[r]{Fairness\\ $F$} \\
\midrule
1.5 & -0.1 & 94.6 & 0.91 & 84.4 & 0.69 \\
2 (default) & 0 & 94.7 & 0.91 & 84.6 & 0.68 \\
3 & +0.4 & 95.3 & 0.86 & 84.8 & 0.68 \\
5 & +0.2 & 94.8 & 0.81 & 85.0 & 0.62 \\
$\infty$ (none) & +1.9 & 95.1 & 0.80 & 87.4 & 0.56 \\
\bottomrule
\end{tabular}
\end{table}

%% file: fig/tab_app_hitrate.tex
\begin{table}[H]
\centering\scriptsize\setlength{\tabcolsep}{6pt}
\caption{Host-cache hit rate versus host-cache size at 0.25 requests/s on $4\times$ H100 with 70\% edits.}
\label{tab:app_hitrate}
\begin{tabular}{lrrrrr}
\toprule
Host cache (GB) & 50 & 100 & 200 & 400 & 800 \\
\midrule
Hit rate (\%) & 32 & 58 & 82 & 93 & 96 \\
\bottomrule
\end{tabular}
\end{table}

%% file: fig/fig7_quality_distilled.tikz
\pgfplotstableread[col sep=comma,comment chars={\#}]{fig/fig7_quality_delta.csv}\qdelta
\colorlet{evalgrey}{black!55}
\pgfplotsset{
  q line/.style={thick,mark size=1.3pt,error bars/y dir=both,error bars/y explicit,
                 error bars/error bar style={line width=.4pt},error bars/error mark options={rotate=90,mark size=1pt}},
  q kvtc/.style={q line,evalgrey,densely dashed,mark=square*,mark options={solid}},
  q ours/.style={q line,evalgreen,very thick,mark=*},
}
\providecommand{\qseries}{}\renewcommand{\qseries}[1]{%
  \addplot[q kvtc] table[x=ratio_x,y=fg,y error plus expr=\thisrow{fg_hi}-\thisrow{fg},
                         y error minus expr=\thisrow{fg}-\thisrow{fg_lo},col sep=comma,comment chars={\#}]{fig/quality/#1_kvtc.csv};
  \addplot[q ours] table[x=ratio_x,y=fg,y error plus expr=\thisrow{fg_hi}-\thisrow{fg},
                         y error minus expr=\thisrow{fg}-\thisrow{fg_lo},col sep=comma,comment chars={\#}]{fig/quality/#1_ours.csv};
}
\providecommand{\qannot}{}\renewcommand{\qannot}[3]{%
  \pgfplotstableread[col sep=comma,comment chars={\#}]{fig/quality/#1_ours.csv}\qours
  \pgfplotstablegetelem{1}{ratio_x}\of\qours\edef\qx{\pgfplotsretval}%
  \pgfplotstablegetelem{1}{fg}\of\qours\edef\qy{\pgfplotsretval}%
  \pgfplotstablegetelem{#2}{dfg}\of\qdelta\edef\qd{\pgfplotsretval}%
  \node[anchor=north east,font=\sffamily\fontsize{6}{7}\selectfont,evalgreen!75!black,inner sep=1pt]
    at (rel axis cs:0.97,0.95) {$\pgfmathprintnumber[fixed,zerofill,precision=1,showpos]{\qd}$\,dB};
}
\begin{tikzpicture}
\begin{groupplot}[evalaxis,
  group style={group size=3 by 1,horizontal sep=0.17in,ylabels at=edge left},
  scale only axis,width=0.80in,height=0.57in,
  xmode=log,log ticks with fixed point,xtick pos=left,ytick pos=left,
  ymin=18,ymax=44,ytick={20,30,40},
  ylabel={PSNR (dB)},
  xlabel={Compression ratio},
  every axis title/.append style={at={(0.5,0)},anchor=north,yshift=-26pt},
  legend style={at={(0.5,1.0)},anchor=south,inner ysep=1pt,legend columns=2,column sep=6pt},
  legend image post style={mark size=1.3pt}]
\nextgroupplot[title={(a) 4 steps},xmin=2.05,xmax=8.6,xtick={2.5,4.5,7},xticklabels={2.5$\times$,4.5$\times$,7$\times$}]
\qseries{distilled_s4}\qannot{distilled_s4}{1}{south west}
\nextgroupplot[title={(b) 6 steps},xmin=2.05,xmax=8.6,xtick={2.5,4.5,7},xticklabels={2.5$\times$,4.5$\times$,7$\times$}]
\qseries{distilled_s6}\qannot{distilled_s6}{4}{south west}
\addlegendentry{KVTC}
\addlegendentry{TileCache}
\nextgroupplot[title={(c) 8 steps},xmin=2.05,xmax=8.6,xtick={2.5,4.5,7},xticklabels={2.5$\times$,4.5$\times$,7$\times$}]
\qseries{distilled_s8}\qannot{distilled_s8}{7}{south west}
\end{groupplot}
\end{tikzpicture}